\documentclass[journal]{IEEEtran}
\usepackage[caption=false,font=footnotesize]{subfig}
\usepackage{cite}
\usepackage{amsmath,amssymb,amsthm}
\usepackage{mathtools}
\usepackage{balance}
\usepackage{graphicx}
\usepackage{tikz}
\usepackage{xcolor}
\usepackage{float}
\usepackage{placeins}
\usepackage{url}
\usepackage{bm}
\usepackage{stfloats}
\usepackage{microtype}
\usepackage[hidelinks]{hyperref}

\newcommand{\RB}[2]{\hat{H}_n^{(#1)}(#2)}
\newcommand{\RBp}[2]{\hat{H}_n^{(#1)\prime}(#2)}

\newtheorem{theorem}{Theorem}
\newtheorem{conjecture}{Conjecture}

\theoremstyle{remark}

\theoremstyle{plain}

\begin{document}
\allowdisplaybreaks

\title{Broadband Absorption Bounds for Coated PEC Spheres: Curvature, Angular Momentum, and Localized Sources}

\author{{Yoav~Kriman,
Amir~Shlivinski,
and~Yakir~Hadad}

\thanks{Y. Kriman and Y. Hadad are with the School of Electrical Engineering, Tel Aviv University, Tel Aviv 6997801, Israel.}%
\thanks{A. Shlivinski is with the School of Electrical and Computer Engineering, Ben-Gurion University of the Negev, Beer-Sheva 84105, Israel.}%
\thanks{Y. Hadad is also with the Department of Electrical and Computer Engineering and the ElectroScience Laboratory, The Ohio State University, Columbus, OH 43212, USA (e-mail: yakirhad@tauex.tau.ac.il).}%
}

\maketitle

\begin{abstract}
We investigate fundamental limits on broadband electromagnetic
absorption by a material-coated perfectly electrically conducting
(PEC) sphere. Exact reflection coefficients are derived for the
transverse-electric (TE) and transverse-magnetic (TM) spherical
modes. Their long-wavelength asymptotics, combined with causality
and contour integration in the complex wavelength plane, yield
sum rules relating the wavelength-integrated modal reflection to
the static material parameters, coating thickness, curvature, and
complex reflection zeros. For the dipolar modes ($n=1$), known
passive spherical-wave sum rules establish analytically a
material-dependent lower bound on the weighted logarithmic
reflection integral. Extension to higher-order modes requires a
nonnegativity condition on higher odd moments of the reflection
zeros; we formulate this condition as a conjecture and provide
extensive Monte-Carlo evidence for passive dispersive coatings.
The resulting bounds establish a per-channel absorption--bandwidth
tradeoff and reveal the spherical-harmonic order as a discrete
transverse-wavenumber variable, recovering the familiar planar
attenuation factor in the locally planar limit. Finally, we show
how the modal bounds can be combined for a localized external
source. For a radial electric dipole outside the sphere, the
near-field excitation is decomposed into TM spherical channels,
leading to a source-weighted broadband absorption bound. As the
source approaches the surface, progressively higher angular
momenta contribute, directly linking near-field spatial
localization to the hierarchy of modal absorption constraints.
\end{abstract}


\section{Introduction}

\IEEEPARstart{E}{lectromagnetic} absorption over a wide bandwidth is of considerable interest in applications such as radar cross-section reduction and stealth technology
\cite{ruck,choi2015,bai2015}, electromagnetic compatibility \cite{holloway,khoshniat}, and compact absorber design \cite{landy,radi,lin2023}.
In a passive structure, however, strong absorption cannot be maintained over an arbitrarily wide band once the available size and the material response are constrained. This limitation stems from the analytic properties that causality and passivity impose on the reflection coefficient, and it underlies the classical broadband matching bounds of Bode and Fano \cite{bode,fano}. For the canonical planar absorber -- a PEC plane covered by a material layer \cite{dallenbach} -- Rozanov translated these analytic constraints into the well-known thickness--bandwidth limit \cite{rozanov}, which has since been refined and extended within the linear time-invariant setting \cite{firestein2023}, also for broad angle incidence \cite{firestein2024}, applied to PEC-backed antenna arrays \cite{doane2013,doane2014}, and generalized to metasurface reflectors and related structures \cite{abdelrahman,gustafsson,schab}.

For \emph{finite} scatterers, the natural analog of the planar reflection coefficient is the modal reflection coefficient of the vector spherical waves, i.e., the diagonal of the scattering matrix in the spherical-wave basis. Bernland, Gustafsson, and Nordebo derived integral sum rules and bandwidth limitations for these coefficients for an arbitrary passive scatterer enclosed in a sphere \cite{Bernland2011VSW}, within the general framework of sum rules for passive systems \cite{Bernland2011SumRules}. Those results employ a fixed, mode-independent spectral weight and yield bounds governed by the circumscribing radius. What is missing is the spherical counterpart of the \emph{Rozanov-type} statement: a per-channel bound, with a spectral weight matched to the long-wavelength order of each angular momentum channel, whose constant is an explicit function of the coating thickness and of the static material parameters, and which reduces to a designer-usable thickness--bandwidth tradeoff. 

This paper develops such bounds for the canonical finite geometry: a PEC sphere of radius $a$ covered by a concentric material coating of thickness $d$, such that the overall radius is $b=a+d$. The main contributions are as follows. First, once the exact TE$_r$ and TM$_r$ modal reflection coefficients are expressed, a weighted contour integration of $\ln R_n^{p}$ in the complex wavelength plane, with the weight $\lambda^{2n}$ matched to the $O\!\left[(k_0b)^{2n+1}\right]$ long-wavelength behavior of the $n$th channel, yields an exact sum rule relating the weighted reflection integral to the static material parameters and to the odd moments of the upper-half-plane reflection zeros. Second, for the dipolar channels ($n=1$) the zero contribution is bounded analytically using the passive spherical-wave sum rules of \cite{Bernland2011VSW,Bernland2011SumRules}, which establish rigorously the absorption bound in this case (Theorem~\ref{thm:dipole}). Third, for $n\ge2$ the required nonnegativity condition on the higher odd zero moments is formulated as an explicit conjecture (Conjecture~\ref{conj:zeros}), supported by Monte Carlo evidence over passive dispersive coatings, as well as by additional physical arguments for the structure of the bound; we also show why the dipolar proof technique cannot extend beyond $n=1$, which makes the remaining conjecture ($n\ge2$) an open problem. Finally, the bound constants admit a transparent geometrical interpretation in which the spherical-harmonic order plays the role of a discrete transverse momentum, and the quasistatic coupling factor $(a/b)^{2n+1}$ approaches the familiar planar attenuation $e^{-2k_td}$ in the locally planar limit.

The remainder of the paper is organized as follows. Section~\ref{sec:formulation} derives the exact modal reflection coefficients and their long-wavelength asymptotics. Section~\ref{sec:bounds} develops the weighted sum rule, states the dipolar theorem and the higher-order conjecture, and gives the resulting absorption bounds together with their Rozanov-type corollary. Section~\ref{sec:numerics} reports the Monte Carlo studies. Section~\ref{sec:geometrical_interpretation} discusses the geometrical interpretation and the connection to planar bounds. Section~\ref{sec:finite_source} demonstrates the application of the theory to absorption by localized sources, and Section~\ref{sec:conclusion} concludes.

\section{Modal Reflection Coefficients of a Coated PEC Sphere}
\label{sec:formulation}

\begin{figure}[t]
\centering

\includegraphics[width=0.99\columnwidth]{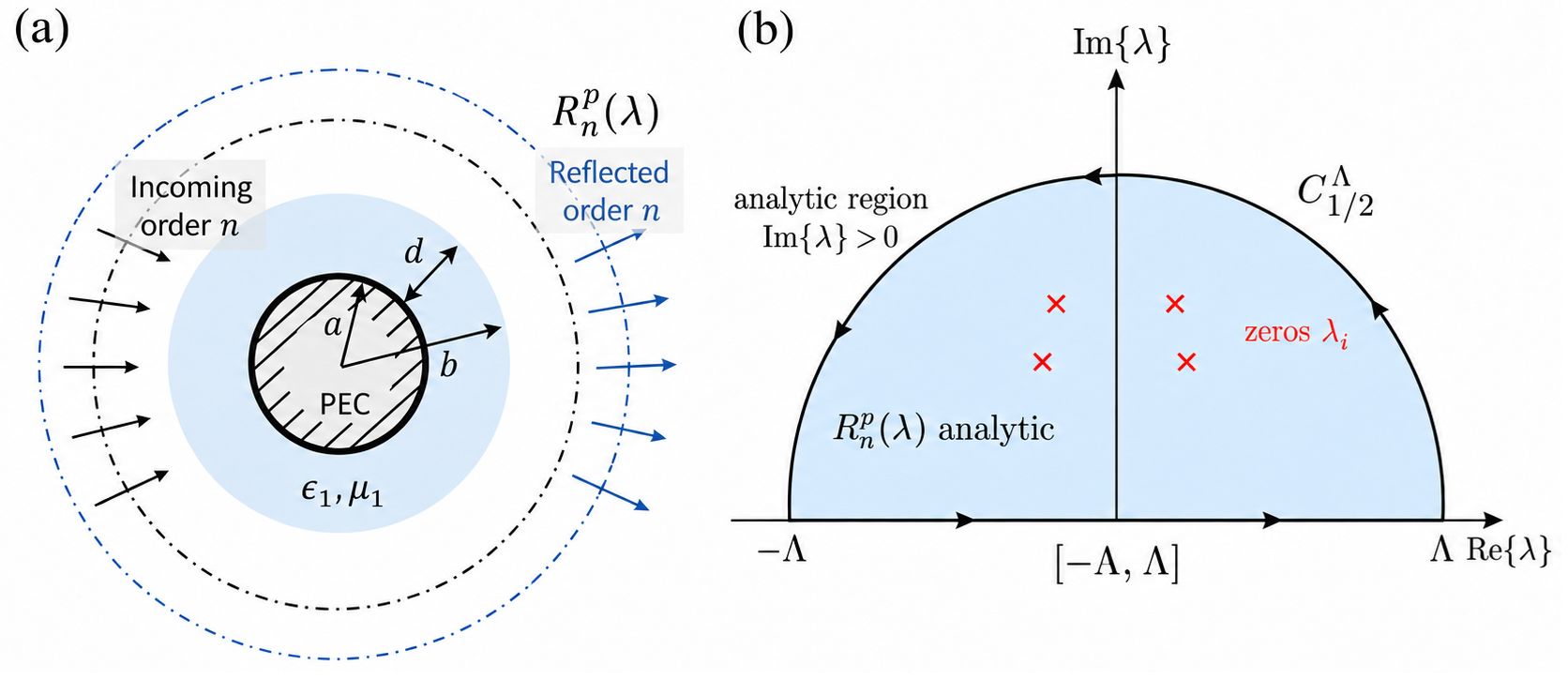}

\caption{
(a) Scattering geometry of a coated PEC sphere. A PEC core of radius
$a$ is surrounded by a material shell of outer radius $b$ and
thickness $d=b-a$, with constitutive parameters $\epsilon_1$ and
$\mu_1$. An incident spherical harmonic of order $n$ generates the
modal reflection coefficient $R_n^p(\lambda)$.
(b) Complex-$\lambda$ contour used in the sum-rule derivation. The
closed contour consists of the real-axis segment $[-\Lambda,\Lambda]$
and the upper semicircular arc $C_{1/2}^{\Lambda}$. The function
$R_n^p(\lambda)$ is analytic in the upper half-plane, with zeros
$\lambda_i$ occurring in symmetry pairs with respect to the imaginary
axis for a real passive system.}
\label{fig:geometry_and_contour}
\label{fig:coated_sphere}

\end{figure}

We consider the coated spherical PEC geometry shown in
Fig.~\ref{fig:coated_sphere}(a), consisting of a PEC core of radius \(a\), a
concentric coating layer with outer radius \(b\), and an exterior free-space
region. Subscript $1$ denotes quantities inside the coating, $a<r<b$, and
subscript $0$ denotes quantities in the exterior region, $r>b$. The coating
parameters are \(\epsilon_1=\epsilon_r\epsilon_0\) and
\(\mu_1=\mu_r\mu_0\), with
\begin{equation*}
 k_0=\omega\sqrt{\mu_0\epsilon_0},
 \qquad
 k_1=\omega\sqrt{\mu_1\epsilon_1}.
\end{equation*}
Throughout, time convention $e^{j\omega t}$ is assumed, and the free-space wavelength $\lambda=2\pi/k_0=2\pi c/\omega$ is used as the spectral parameter.  For a passive coating, causality implies that $R_n^{p}(\lambda)$ defined below is analytic in the lower half of the complex $\omega$ plane and, correspondingly, analytic in the \emph{upper} half of the complex $\lambda$ plane, with $|R_n^{p}(\lambda)|\le1$ \emph{on the real axis} \cite{Bernland2011SumRules,jackson}.

The modal fields are represented using radial vector potentials,
\(\mathbf{A}=\hat{\mathbf r}\Psi_a\) for TM\(_r\) modes and
\(\mathbf{F}=\hat{\mathbf r}\Psi_f\) for TE\(_r\) modes, with radial dependence given by the Riccati-Bessel functions
\begin{equation}
\hat H_n^{(1,2)}(x) = x\,h_n^{(1,2)}(x),
\label{eq:RB_def}
\end{equation}
where $h_n^{(1,2)}$ are the spherical Hankel functions and, with the $e^{j\omega t}$ convention, the superscript $(1)$ $(2)$ corresponds to an inward and outward propagating spherical wave respectively \cite{harrington,balanis}. The incident (incoming) potential of the $n$th harmonic is $\Psi^{\mathrm{inc}}=\hat H_n^{(1)}(k_0 r)\,P_n^{m}(\cos\theta)\,e^{jm\phi}$ \cite{harrington}; since the  problem is degenerate in the azimuthal index, the reflection coefficients below depend on $n$ and on the polarization $p\in\{\mathrm{TE},\mathrm{TM}\}$ only. Source-free electromagnetic fields contain no $n=0$ spherical mode, so the modal index starts at the dipole, $n\ge1$. Primes denote
differentiation with respect to the argument. Boundary conditions on the inner conductor at $r=a$ [$\Psi_f=0$ for TE$_r$; $\partial_r\Psi_a=0$ for TM$_r$] and field continuity at the coating interface $r=b$ yield the TE and TM coating coefficients

\begin{subequations}
\label{eq:Q}

\begin{equation}
\resizebox{0.88\columnwidth}{!}{$\displaystyle
Q_{\mathrm{TE}}=
\frac{
\RBp{2}{k_1 b}\RB{1}{k_1 a}
-
\RBp{1}{k_1 b}\RB{2}{k_1 a}
}{
\RB{2}{k_1 b}\RB{1}{k_1 a}
-
\RB{2}{k_1 a}\RB{1}{k_1 b}
}
\sqrt{\frac{\epsilon_r}{\mu_r}}
$}
\label{eq:QTE}
\end{equation}

\begin{equation}
\resizebox{0.88\columnwidth}{!}{$\displaystyle
Q_{\mathrm{TM}}=
\frac{
\RBp{2}{k_1 b}\RBp{1}{k_1 a}
-
\RBp{2}{k_1 a}\RBp{1}{k_1 b}
}{
\RB{2}{k_1 b}\RBp{1}{k_1 a}
-
\RB{1}{k_1 b}\RBp{2}{k_1 a}
}
\sqrt{\frac{\mu_r}{\epsilon_r}}
$}
\label{eq:QTM}
\end{equation}

\end{subequations}

For \(p\in\{\mathrm{TE},\mathrm{TM}\}\), let \(Q_p\) denote the corresponding
coating coefficient in Eqs.~\eqref{eq:QTE} and~\eqref{eq:QTM}. The modal
reflection coefficient of the \(n\)-th spherical harmonic is
{
\begin{equation}
R_n^{p}(\lambda)=
\frac{
\RBp{1}{k_0b}-Q_p \RB{1}{k_0b}
}{
Q_p \RB{2}{k_0b}-\RBp{2}{k_0b}
}.
\label{eq:exact_R}
\end{equation}
}%

Next, we consider the long-wavelength behavior of Eq.~(\ref{eq:exact_R}).  For the sake of brevity, we define and use below
\begin{equation*}
x \equiv k_0b=\frac{2\pi b}{\lambda}\ll1.
\end{equation*}
Using known Riccati--Bessel small argument expansions (Appendix~\ref{app:hankel_expansions}), we obtain
\begin{equation}
\ln R_n^p(\lambda)
\sim
j\left(A_n-B_n^p\right)\lambda^{-(2n+1)}.
\label{eq:R_full_asymptotic_n}
\end{equation}
with
\begin{equation}
A_n
=
\frac{2(2\pi b)^{2n+1}}
{(2n+1)!!(2n-1)!!}
\label{eq:An}
\end{equation}
and
\begin{equation}
B_n^p
=
A_n\frac{2n+1}{C_p+n}.
\label{eq:Bn}
\end{equation}
The material-dependent real-valued constants (evaluated at the static values of $\epsilon_r$, $\mu_r$)
read,
\begin{equation}
\begin{aligned}
C_{\mathrm{TM}}
&=
\frac{1}{\epsilon_r}
\frac{n(n+1)\left[1-(a/b)^{2n+1}\right]}
{n+(n+1)(a/b)^{2n+1}},
\\
C_{\mathrm{TE}}
&=
\frac{1}{\mu_r}
\frac{(n+1)+n(a/b)^{2n+1}}
{1-(a/b)^{2n+1}}.
\end{aligned}
\label{eq:Cm}
\end{equation}
Both constants contain the quasistatic coupling factor $(a/b)^{2n+1}$, whose geometrical origin is discussed in Section~\ref{sec:geometrical_interpretation}.
{Throughout, $\epsilon_r$ and $\mu_r$ denote the static values $\epsilon_r(0)$
and $\mu_r(0)$ giving that in the long-wavelength limit the coated sphere reduces to an electrostatic (TM) or magnetostatic (TE) core--shell problem, and $C_p$ is a
problem's polarizability factor. Dispersion therefore does not enter these coefficients, it contributes at higher order terms in $1/\lambda$ leaving $A_n$, $B_n^p$ independent of $\lambda$.}
\section{Weighted Sum Rules and Modal Absorption Bounds}
\label{sec:bounds}

The analytic properties of the reflection coefficient in passive systems impose fundamental integral constraints on broadband absorption \cite{bode,fano,rozanov,firestein2023}. Although \(R_n^p(\lambda)\) is
analytic in the upper half \(\lambda\) plane, \(\ln R_n^p(\lambda)\) may
have branch points originating from the zeros of \(R_n^p\) (see Fig.~\ref{fig:coated_sphere}(b)). To account for these, we introduce the Blaschke (all-pass) product that removes the upper-half-plane zeros of $R_n^p(\lambda)$
\cite{bode,fano},
\begin{equation}
H_{\mathrm{AP}}(\lambda)
=
\prod_{i=1}^{N}
\frac{\lambda-\lambda_i^*}
{\lambda-\lambda_i},
\label{eq:HAP}
\end{equation}
where $\lambda_i$ are the zeros of $R_n^p(\lambda)$ \emph{in the upper half plane}; by the conjugate symmetry of a real passive response they occur in pairs $(\lambda_i,-\lambda_i^*)$. Here $N$ is assumed finite; more generally, the derivation requires the zero moments appearing below to be absolutely convergent; see the corresponding Blaschke-product conditions in \cite{Bernland2011SumRules}.
Using \eqref{eq:exact_R} and \eqref{eq:HAP}, we define the phase-reduced reflection coefficient
\begin{equation}
\widetilde{R}_n^p(\lambda)
=
R_n^p(\lambda)H_{\mathrm{AP}}(\lambda).
\label{eq:R_tilde}
\end{equation}
The phase-reduced coefficient in Eq.~\eqref{eq:R_tilde}
has no zeros in the upper half $\lambda$ plane, making
$\ln\widetilde{R}_n^p(\lambda)$ analytic in that domain.
Let $W(\lambda)$ be a polynomial weight, real on the real axis. Applying Cauchy's theorem to $W\ln\widetilde R_n^p$ over the closed contour consisting of the segment $[-\Lambda,\Lambda]$ of the real axis and the upper semicircular arc $C_{1/2}^{\Lambda}$ of radius $\Lambda$, taking the real part, and letting $\Lambda\to\infty$ gives (see Fig.~\ref{fig:coated_sphere}(b))
\begin{equation}
\begin{aligned}
\int_{-\infty}^{\infty}
W(\lambda)\ln\bigl|\widetilde{R}_{n}^p(\lambda)\bigr|\,d\lambda
&=
-\lim_{\Lambda\to\infty}\Re\!\int_{C_{1/2}^{\Lambda}}
W\ln R_n^p\,d\lambda
\\
&\phantom{=}
-\lim_{\Lambda\to\infty}\Re\!\int_{C_{1/2}^{\Lambda}}
W\ln H_{\mathrm{AP}}\,d\lambda.
\end{aligned}
\label{eq:contour_identity}
\end{equation}
On the left, $\ln|\widetilde R_n^p|=\ln|R_n^p|$ by Eq.~\eqref{eq:R_tilde} since the all-pass $|H_{\rm AP}|=1$ on the real axis -- not affecting the reflection magnitude. Taking the real part before the limit is essential, since the terms that grow with $\Lambda$ are purely imaginary and vanish upon taking the real part. (Appendix~\ref{app:all_pass}).

For \(n\geq1\), the weight is chosen as $W(\lambda)=\lambda^{2n}$,  for which the leading long-wavelength term of $\ln R_n^p$ produces a finite, nonzero arc contribution.  Since
\(
\left(A_n-B_n^p\right)\lambda^{-(2n+1)}
\rightarrow0
\)
as \(\lvert\lambda\rvert\rightarrow\infty\), the approximation
\(\ln(1+z)\sim z\) gives $\ln R_n^p(\lambda)
\sim
j\left(A_n-B_n^p\right)\lambda^{-(2n+1)}$. Then, integration along the upper semicircular arc reads
\begin{equation}
\lim_{\Lambda\to\infty}\int_{C_{1/2}^{\Lambda}}
\lambda^{2n}\ln R_n^p(\lambda)\,d\lambda
=
\pi B_n^p-\pi A_n.
\label{eq:R_full_arc}
\end{equation}
The all-pass contribution, derived in
Appendix~\ref{app:all_pass}, is
\begin{equation}
\resizebox{0.88\columnwidth}{!}{$\displaystyle
\lim_{\Lambda\to\infty}\!\Re\!\int_{C_{1/2}^{\Lambda}}
\lambda^{2n}\ln H_{\mathrm{AP}}(\lambda)\,d\lambda
=
-\frac{2\pi}{2n+1}
\sum_{i=1}^{N}
\Im\!\left\{\lambda_i^{2n+1}\right\}.
$}
\label{eq:HAP_arc}
\end{equation}

Substituting Eqs.~\eqref{eq:R_full_arc} and~\eqref{eq:HAP_arc} into
Eq.~\eqref{eq:contour_identity} yields the exact sum rule
\begin{equation}
\begin{aligned}
\int_{-\infty}^{\infty}
\lambda^{2n}\ln|R_n^p(\lambda)|\,d\lambda
&=
-\pi B_n^p+\pi A_n
\\
&\quad+
\frac{2\pi}{2n+1}
\sum_{i=1}^{N}
\Im\!\left\{\lambda_i^{2n+1}\right\},
\end{aligned}
\label{eq:integral_n}
\end{equation}
which relates the weighted logarithmic reflection integral to the static material parameters (through $B_n^p$) and to the odd moments of the upper-half-plane reflection zeros.

The sign of the zero moment $\sum_i\Im\{\lambda_i^{2n+1}\}$ is not fixed a priori: writing $\lambda_i=2\pi b\,(x_i+jy_i)$ with $y_i>0$, the individual terms $\Im\{\lambda_i^{2n+1}\}$ can be of either sign depending on the zero locations. The usefulness of Eq.~\eqref{eq:integral_n} as a bound therefore rests on the nonnegativity of the combined geometric-plus-zero contribution:
\begin{equation}
\mathcal{G}_n^{(p)}=\pi A_n
+
\frac{2\pi}{2n+1}
\sum_{i=1}^{N}
\Im\!\left\{
\lambda_i^{2n+1}
\right\}
\geq 0,
\qquad n\geq 1.
\label{eq:nonnegative_combined_higher_modes}
\end{equation}
Since $A_n>0$, any negative contribution can only originate from the reflection zeros. For the dipolar channels this condition is not an assumption but a theorem:

\begin{theorem}[Dipolar modes]
\label{thm:dipole}
For any passive coating, the TE and TM reflection coefficients of the $n=1$ spherical channels satisfy Eq.~\eqref{eq:nonnegative_combined_higher_modes}. 
\end{theorem}
The proof, given in Appendix~\ref{app:dipole_proof}, bounds the zero moment using the passive spherical wave sum rules of Bernland \emph{et al.} \cite{Bernland2011VSW,Bernland2011SumRules}, and holds in fact for \emph{any} passive scatterer circumscribed by the radius $b$.

For $n\ge2$, the elementary inequality used at $n=1$ demonstrably fails: the dipolar proof in Appendix~\ref{app:dipole_proof}  rests on $\Im\{z^3\}\ge-\Im\{z\}^3$ for $\Im\{z\}>0$, whereas already for  $n=2$ a similar inequality with $z^5$ doesn't hold. For example,  $z=1+j$ gives $\Im\{z^5\}=-4\not\ge-\Im\{z\}^5=-1$. Any proof for $n\ge2$ must therefore exploit \emph{collective constraints on the zero set} rather than per-zero estimates. Thus, for $n\ge2$, lacking such a rigorous proof, we are left with a conjecture supported below by extensive Monte-Carlo simulations, as well as by the physical intuition that the bound should be a monomial term depending simultaneously on geometry and material at $\lambda\rightarrow\infty$, like $B_n^p$. Strictly speaking, the conjecture below is a mathematical puzzle to be solved.

\begin{conjecture}[Higher-order zero moments]
\label{conj:zeros}
Let $\{z_i\}=\{\lambda_i/(2\pi b)\}$ be the upper-half-plane zeros of $R_n^p(\lambda)$, normalized by $2\pi b$, for a coated PEC sphere with any passive coating. Then, for every $n\ge2$ and $p\in\{\mathrm{TE},\mathrm{TM}\}$,
\begin{equation}
\sum_i \Im\!\left\{z_i^{\,2n+1}\right\}\;\ge\;-\frac{1}{\left[(2n-1)!!\right]^{2}}.
\end{equation}
\end{conjecture}
Conjecture~\ref{conj:zeros}  is a rearrangement of the nonnegativity requirement stated in Eq.~(\ref{eq:nonnegative_combined_higher_modes}) using Eq.~(\ref{eq:An}). It is supported by the Monte-Carlo evidence of Section~\ref{sec:numerics}. 
Note that on the left-hand side we should sum over all the zeros of $R^p_{n}$, and the negative term of the right-hand side originates from $A_n$, which can readily be interpreted as the \emph{sum over the moments of all the zeros of $h_n^{(1)}(k_0 b)/h_n^{(2)}(k_0 b)$} of a pure delay all-pass spherical filter (also, up to a sign, the TE reflection coefficient of a bare PEC sphere). Thus Eq.~\eqref{eq:nonnegative_combined_higher_modes}
indeed implies that the sum over the moments of the zero set of \emph{all} the zeros of $R^p_n(\lambda)$ is bounded from below, as physically expected, anticipating the existence of a bound.  
Under Theorem~\ref{thm:dipole} for $n=1$,  and Conjecture~\ref{conj:zeros} for $n\ge2$, Eq.~\eqref{eq:integral_n} gives
\begin{equation}
I_n^p=\int_{0}^{\infty}
\lambda^{2n}\ln|R_n^p(\lambda)|\,d\lambda
\geq
-\frac{\pi}{2} B_n^p,
\qquad n\geq1,
\label{eq:bound_n}
\end{equation}
where we utilized the conjugate symmetry of a passive real material response
\cite{jackson},
\begin{equation*}
|R_n^p(-\lambda)|=|R_n^p(\lambda)|,
\end{equation*}
to reduce the integration path to $[0,\infty)$.

Note that although the exact reflection coefficient generally depends on both
\(\epsilon_r\) and \(\mu_r\), the TE bound does not contain
\(\epsilon_r\) explicitly, while the TM bound does not contain
\(\mu_r\) explicitly. This selective dependence is used in the following Monte Carlo
simulations.

\subsection{A Rozanov-Type Absorption--Bandwidth Tradeoff}
The engineering content of Eq.~\eqref{eq:bound_n} parallels Rozanov's thickness--bandwidth limit \cite{rozanov}. Suppose a design requires $|R_n^p(\lambda)|\le\rho_0<1$ over the band $\lambda\in[\lambda_1,\lambda_2]$. Since $\ln|R_n^p|\le0$ everywhere on the real axis, restricting the (halved) bound to the design band gives
\begin{equation}
\ln\frac{1}{\rho_0}\,
\frac{\lambda_2^{2n+1}-\lambda_1^{2n+1}}{2n+1}
\;\le\;
\frac{\pi}{2}\,B_n^p .
\label{eq:rozanov_corollary}
\end{equation}
Equation~\eqref{eq:rozanov_corollary} therefore establishes an absorption--bandwidth tradeoff for each spherical channel, governed by the $(2n+1)$th power of the band-edge wavelengths. Consequently, the constraint is weighted increasingly toward the long-wavelength edge.

As an illustration, for the dipolar TM channel with $b/a=1.1$,
$a=1~\mathrm{cm}$, and a coating of static permittivity
$\epsilon_s=10$, Eq.~(16) limits a 20-dB modal reflectivity
reduction ($\rho_0=0.1$) to bands satisfying
$\lambda_2^3-\lambda_1^3 \lesssim 1.325\times10^{-3}\mathrm{m}^3$.
For example, this corresponds to a bandwidth of approximately
$16.4~\mathrm{MHz}$ around $f_c=1~\mathrm{GHz}$, or approximately
$259~\mathrm{MHz}$ around $f_c=2~\mathrm{GHz}$.
For the quadrupolar TM channel, $n=2$, under the same geometrical
and material conditions, the corresponding constraint is $\lambda_2^5-\lambda_1^5
\lesssim 5.806\times10^{-7}\ \mathrm{m}^5$
This translates to a bandwidth of only about
$48~\mathrm{kHz}$ around $f_c=1~\mathrm{GHz}$ and
$3.07~\mathrm{MHz}$ around $f_c=2~\mathrm{GHz}$.
The much stronger bandwidth restriction for $n=2$ illustrates
the rapidly increasing long-wavelength penalty associated with
higher-order spherical channels.

\section{Monte Carlo Bound Tests}
\label{sec:numerics}

The validity of the nonnegativity condition stated in Eq.~\eqref{eq:nonnegative_combined_higher_modes}, or equivalently, Conjecture~\ref{conj:zeros} -- stating the bounds are examined below using a series of Monte-Carlo simulations.
For \(n=1,2,3\), a set of 500 random material realizations were generated at each
swept parameter value to test the bound in Eq.~\eqref{eq:bound_n}.
For each realization, the geometrical parameters were fixed at
\(a=1~\mathrm{cm}\) and \(b/a=1.1\), and the positive wavelength integral
\begin{equation}
I_n^p=\int_{0}^{\infty}
\lambda^{2n}\ln|R_n^p(\lambda)|\,d\lambda,
\qquad n=1,2,3,
\end{equation}
was evaluated numerically and compared with the bound $-\pi B_n^p/2$.
For the TE modes, the analytical bound contains \(\mu_r\) but not
\(\epsilon_r\). Therefore,
\(
\mu_r=1,2,\ldots,50
\)
was swept while \(\epsilon_r\) was randomized. Here, $\mu_r$ was taken
as a real constant, while $\epsilon_r$ was modeled using a Lorentz model
\begin{equation}
\label{eq:eps_drude}
\epsilon_r(\omega)
=
\epsilon_\infty
-
\frac{\omega_p^2}
{\omega^2-j\omega\Gamma}
\end{equation}
with the uniformly sampled parameters
\begin{equation*}
\epsilon_\infty\in[1,10^6],
\qquad
\omega_{p,\epsilon},\Gamma_\epsilon
\in[1,10^{14}]~\mathrm{rad/s}.
\end{equation*}
For the TM modes, \(n=1,2,3\), $\epsilon_r$ was swept and taken as a real constant,
\(
\epsilon_r=1,2,\ldots,50,
\)
while the permeability was randomized using the magnetic Lorentz model
\begin{equation}
\label{eq:mu_drude}
\mu_r(\omega)
=
\mu_\infty-
\frac{\omega_{p,\mu}^{2}}
{\omega^2-\omega_{0}^2-j\Gamma_\mu\omega},
\end{equation}
with the uniformly sampled parameters
\begin{equation*}
\mu_\infty\!\in\![1,10^6],
\mbox{ }
\omega_{p.\mu},\Gamma_{\mu}\!\in\![1,10^{14}]\mbox{rad/s },
\omega_0\!\in\![10^{8},10^{12}]\mbox{rad/s }
\end{equation*}
{Note that the dielectric model in Eq. (\ref{eq:eps_drude}) exhibits a divergent static loss, \(\epsilon_r''\sim\omega_p^2/(\Gamma\omega)\) as \(\omega\to 0\), which has a direct physical interpretation in terms of static conductivity due to free charge carriers. In contrast, such a divergence in the magnetic response has no analogous physical interpretation. Therefore, we use the magnetic model in Eq. (\ref{eq:mu_drude}) that ensures a finite low-frequency response. 
When comparing the two models in Eqs.~(\ref{eq:eps_drude}) and (\ref{eq:mu_drude}) , it is interesting to note that while the TE (TM) bound itself is independent of $\epsilon_r$ ($\mu_r$), the derivation's low-frequency hypotheses should be kept in mind for such materials (cf.\ Section~\ref{sec:bounds}). Nevertheless, the final result remains completely independent of that behavior - even in the presence of singular low-frequency behavior such as in the electric material properties.}

\begin{figure}[!tbp]
    \centering
    \includegraphics[width=\columnwidth]{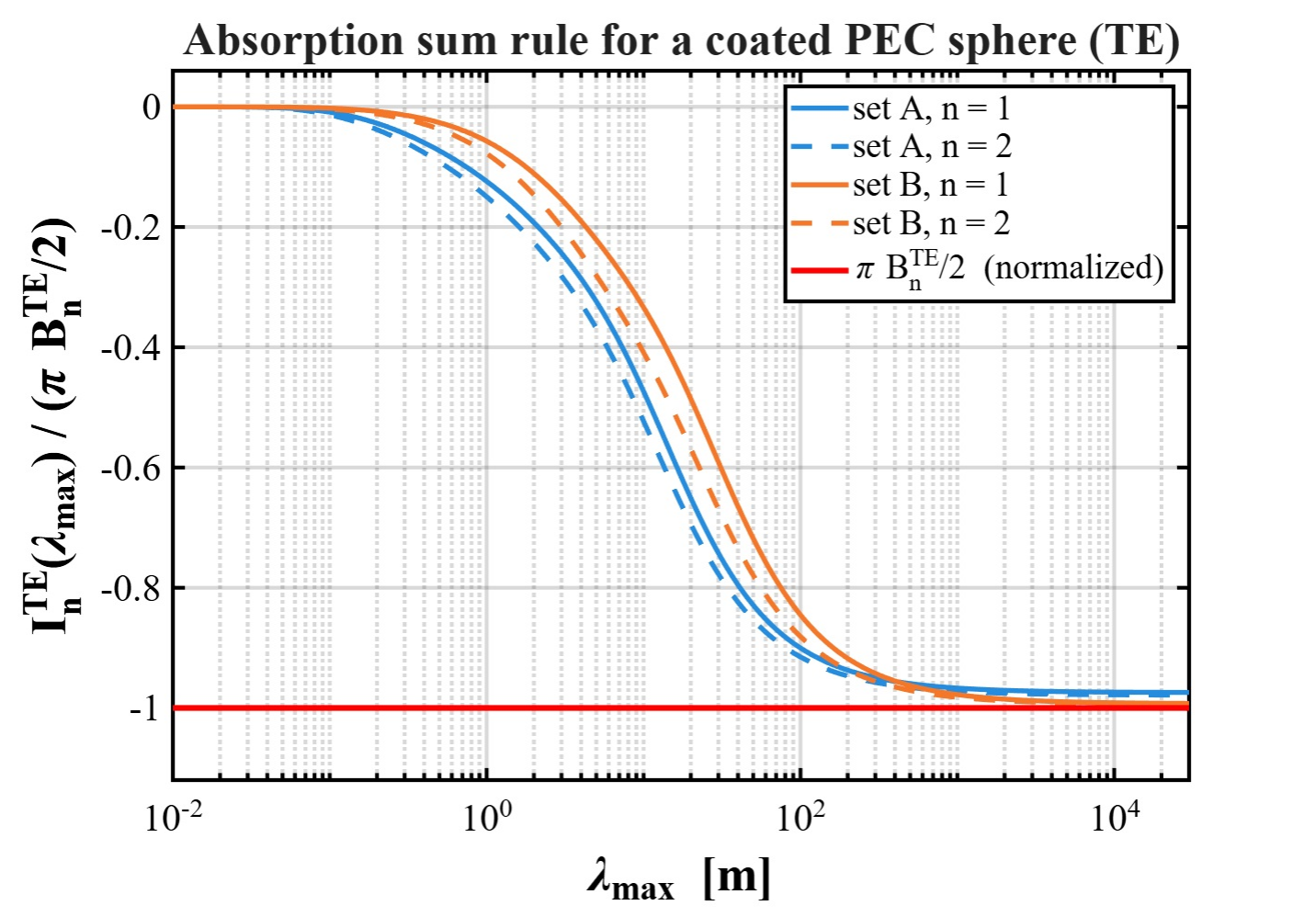}
    \caption{Convergence of the normalized TE absorption integral
    as a function of the upper integration limit $\lambda_{\max}$
    for two sets of material parameters (A and B) and harmonic
    orders $n=1$ and $n=2$. The solid and dashed curves correspond
    to $n=1$ and $n=2$, respectively. The horizontal red line
    represents the normalized analytical bound. All four cases
    exhibit rapid convergence to values close to the analytical
    bound.}
    \label{fig:TE_convergence}
\end{figure}

{
The numerical evaluation of the absorption integral is challenging
due to the rapid oscillations of the integrand arising from the
clustering of zeros near $\lambda=0$. To accurately resolve these
oscillations, a logarithmically scaled, dynamically refined
integration grid was employed. For each harmonic order, the
integration range was independently optimized to ensure convergence
while maintaining sufficient numerical resolution.
Fig.~\ref{fig:TE_convergence} demonstrates the  convergence
of the normalized absorption integral with increasing
$\lambda_{\max}$ for two sets of material parameters in the
TE polarization, considering harmonic orders $n=1$ and $n=2$.
In all four cases, the integral converges rapidly to a value
close to the analytical bound, with only minor variations
observed upon further increasing the integration range.
Similar results for TM case exist but are not shown in for brevity. 
}

\begin{figure*}[!t]
\centering

\begin{minipage}{0.31\textwidth}
    \centering
    \includegraphics[width=\linewidth]{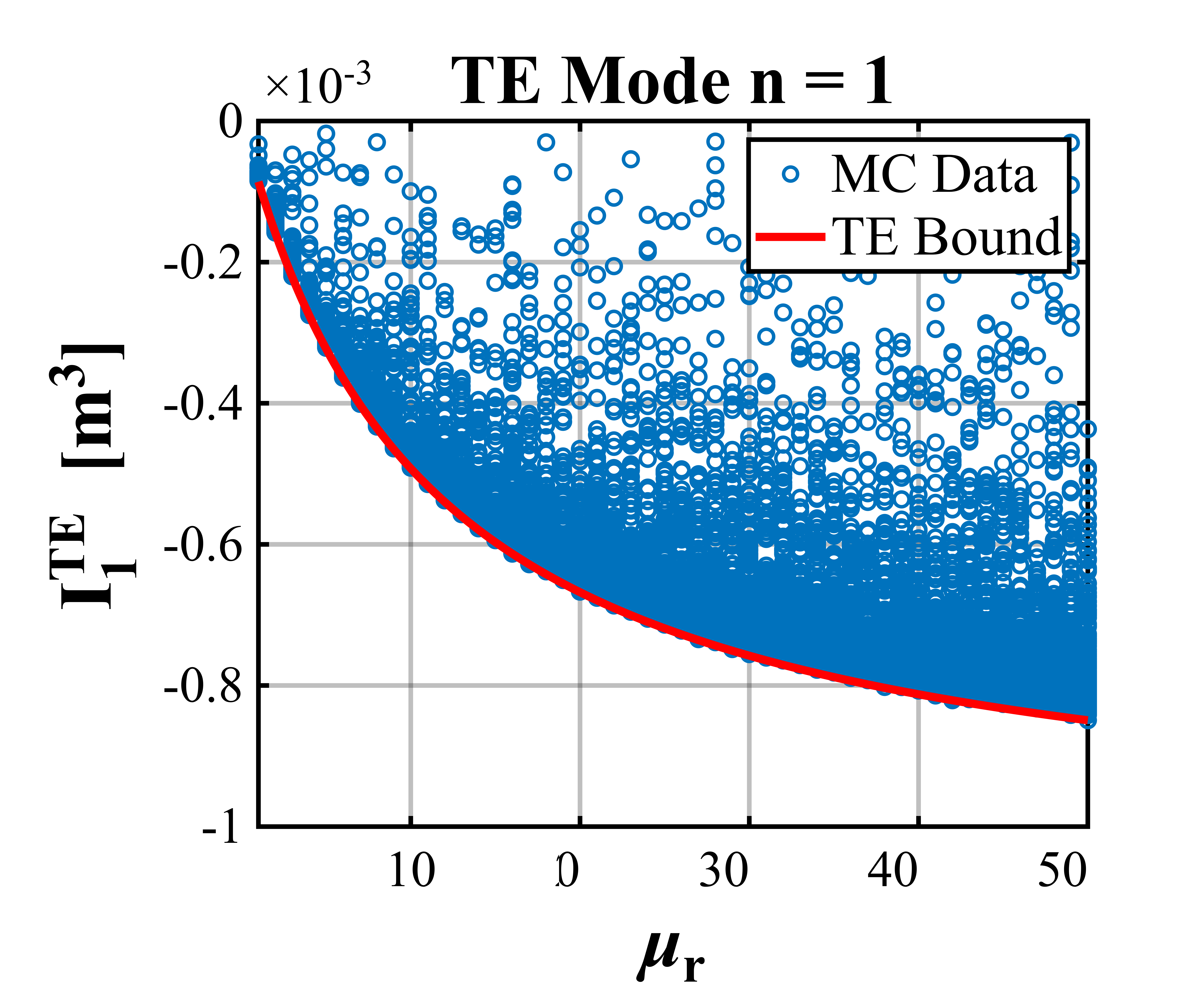}\\[-1mm]
    \hspace{3mm}(a)
\end{minipage}
\hfill
\begin{minipage}{0.31\textwidth}
    \centering
    \includegraphics[width=\linewidth]{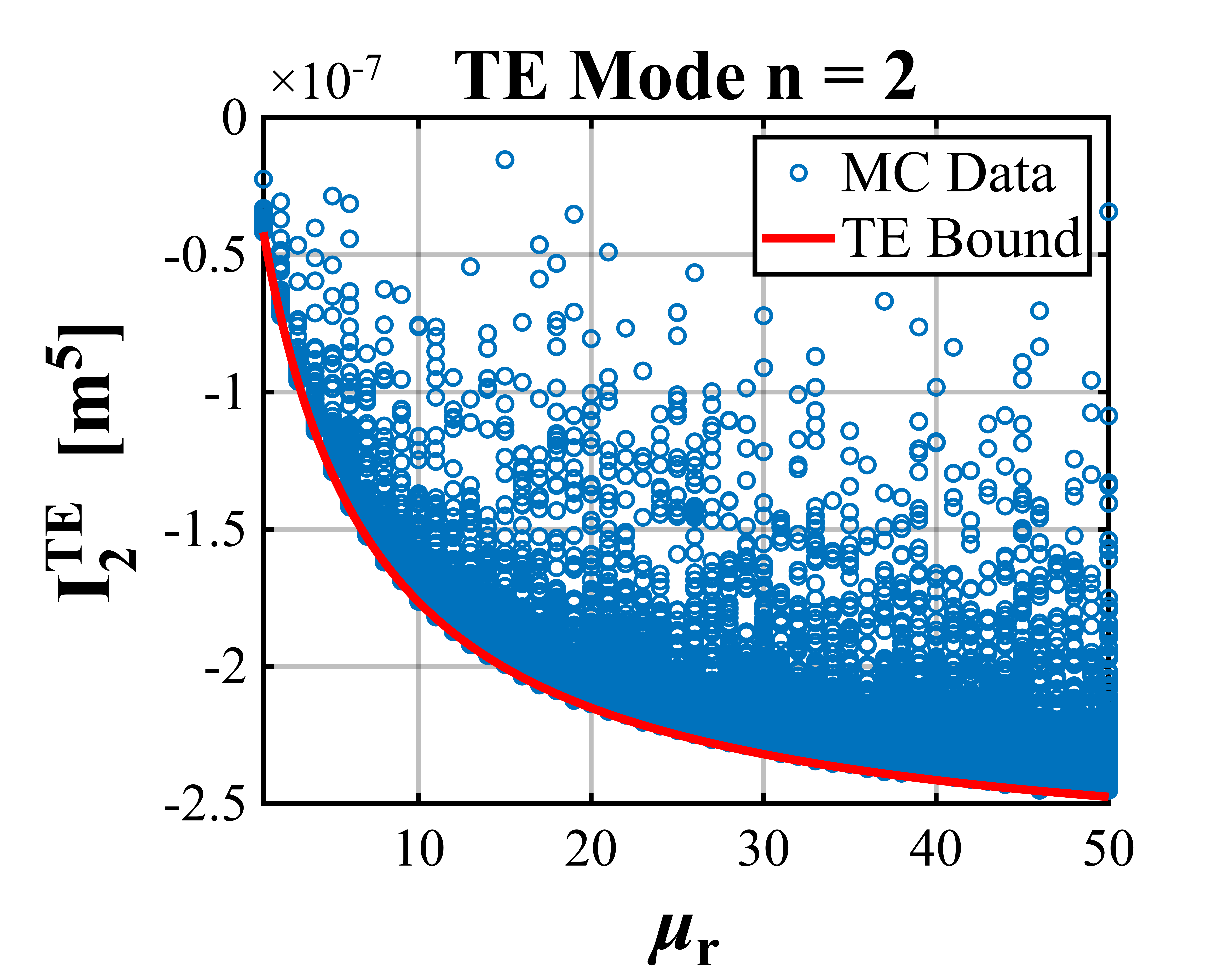}\\[-1mm]
    \hspace{3mm}(b)
\end{minipage}
\hfill
\begin{minipage}{0.31\textwidth}
    \centering
    \includegraphics[width=\linewidth]{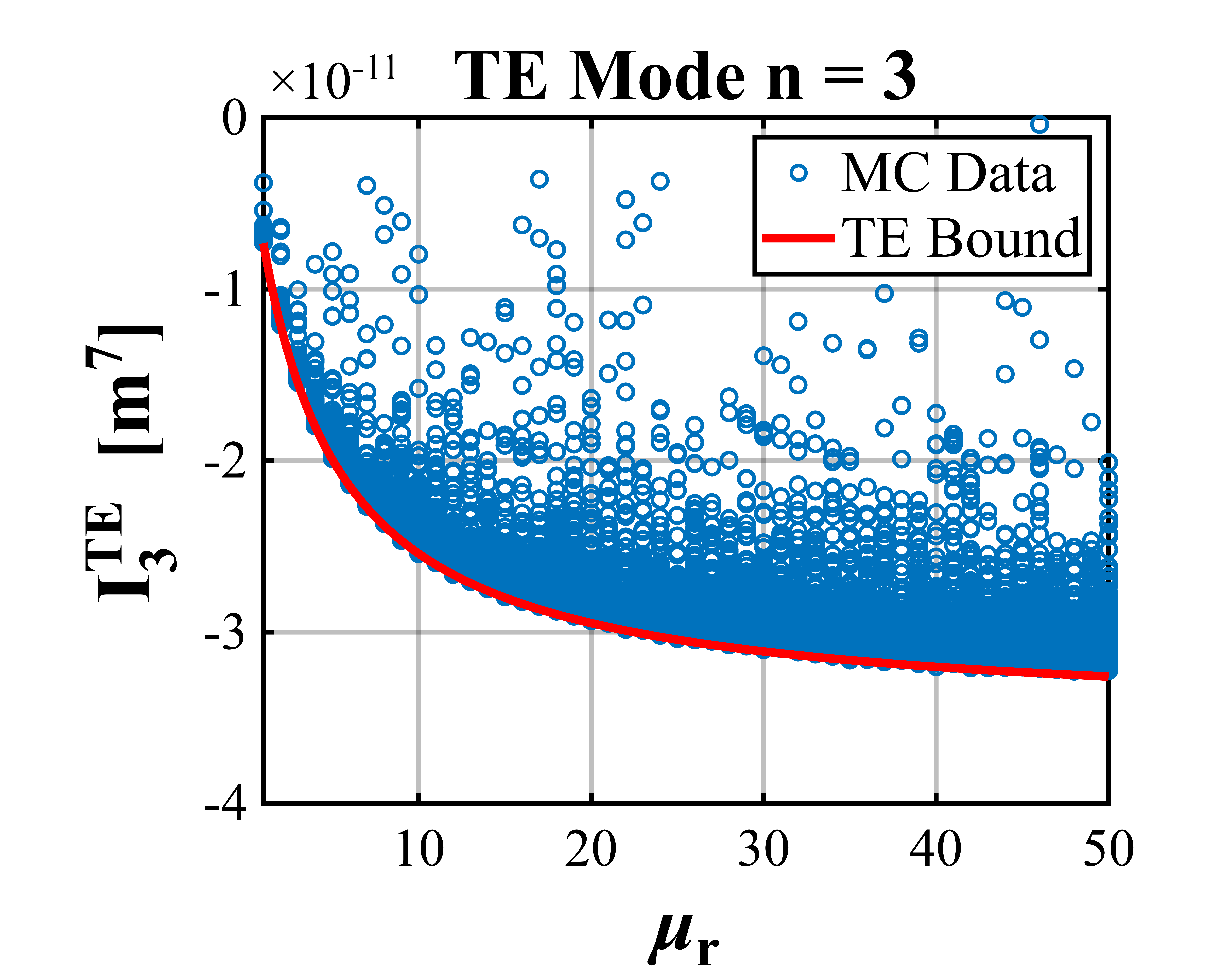}\\[-1mm]
    \hspace{3mm}(c)
\end{minipage}

\vspace{1.5mm}

\begin{minipage}{0.31\textwidth}
    \centering
    \includegraphics[width=\linewidth]{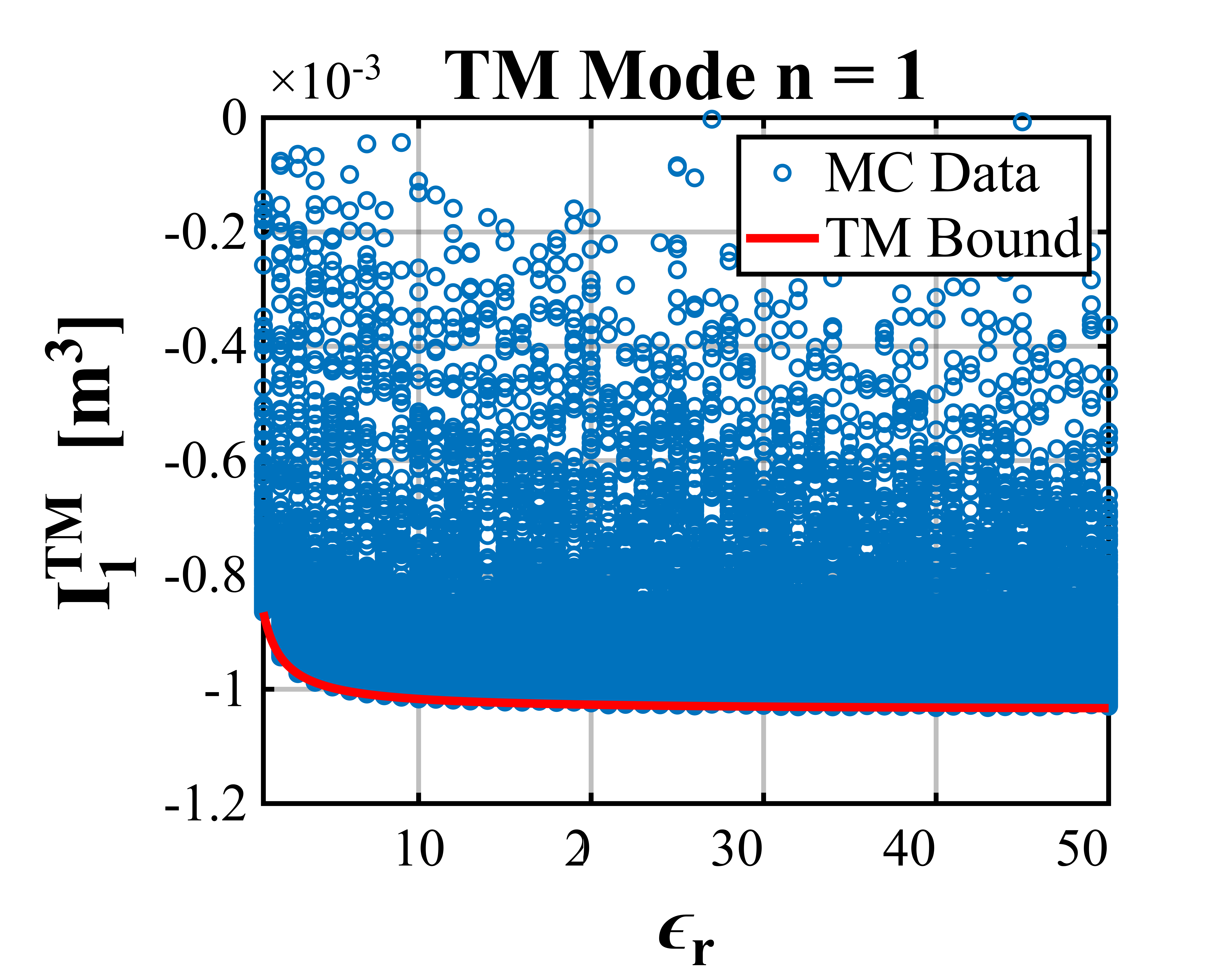}\\[-1mm]
    \hspace{3mm}(d)
\end{minipage}
\hfill
\begin{minipage}{0.31\textwidth}
    \centering
    \includegraphics[width=\linewidth]{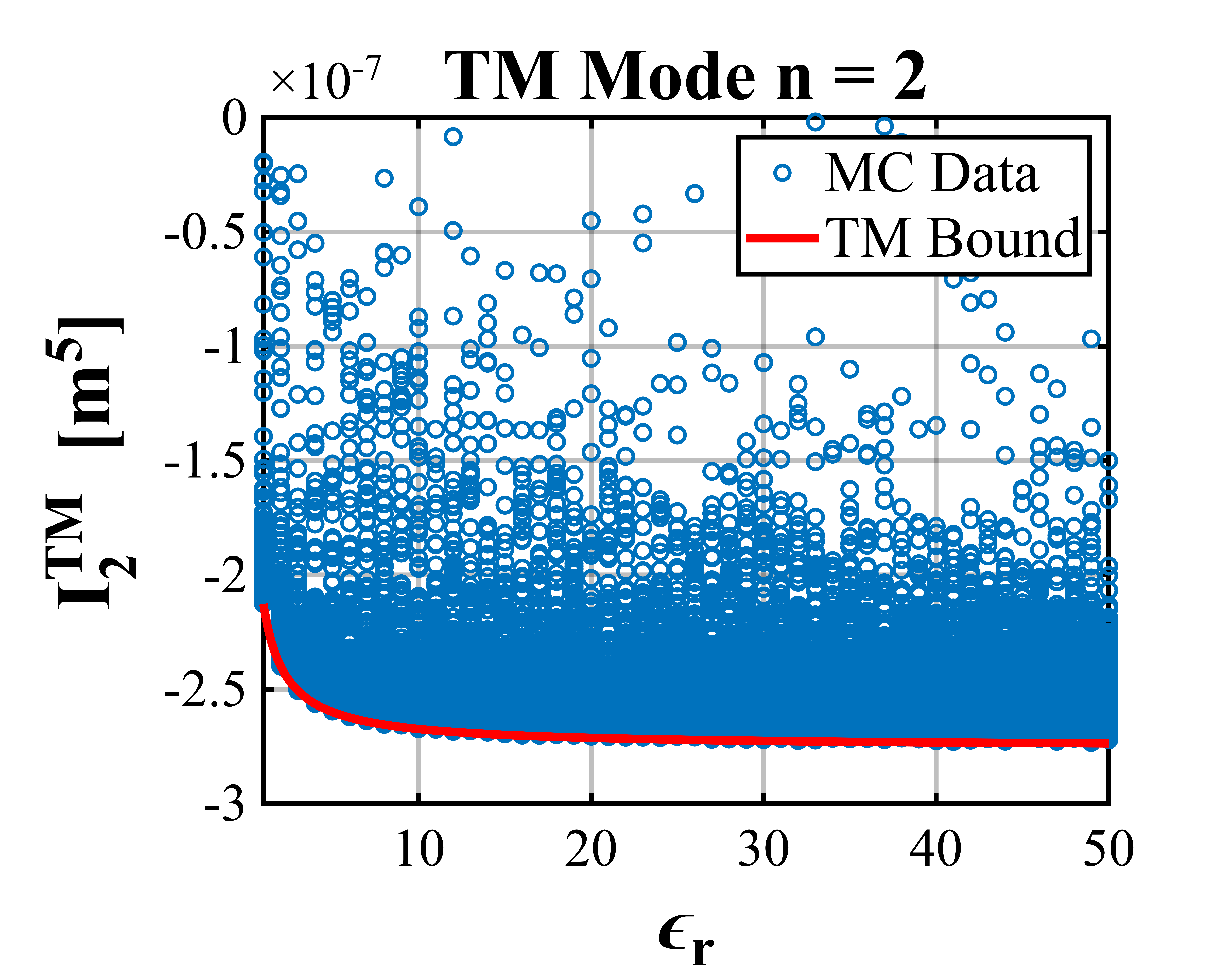}\\[-1mm]
    \hspace{3mm}(e)
\end{minipage}
\hfill
\begin{minipage}{0.31\textwidth}
    \centering
    \includegraphics[width=\linewidth]{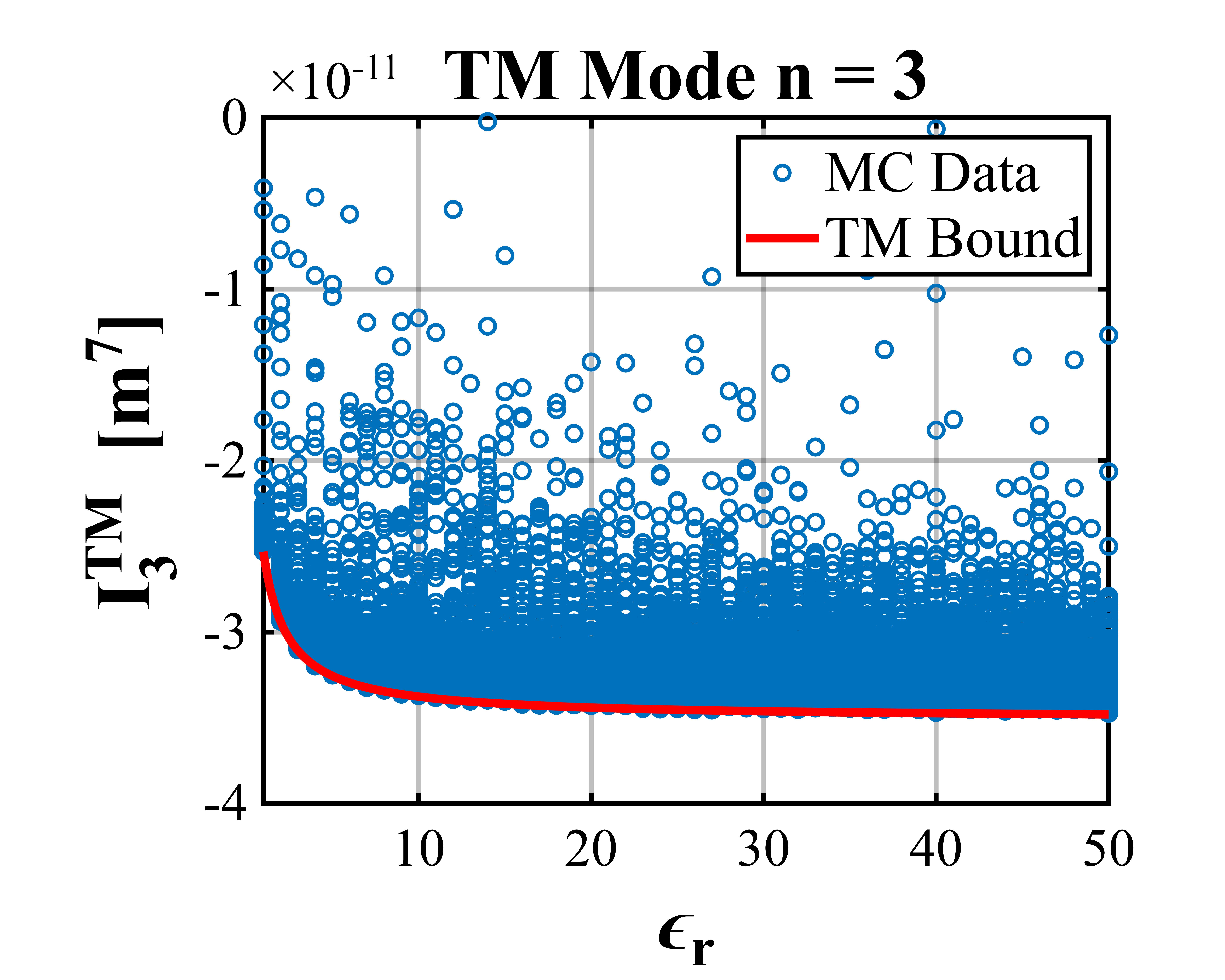}\\[-1mm]
    \hspace{3mm}(f)
\end{minipage}

\vspace{1.5mm}

\begin{minipage}{0.31\textwidth}
    \centering
    \includegraphics[width=\linewidth]{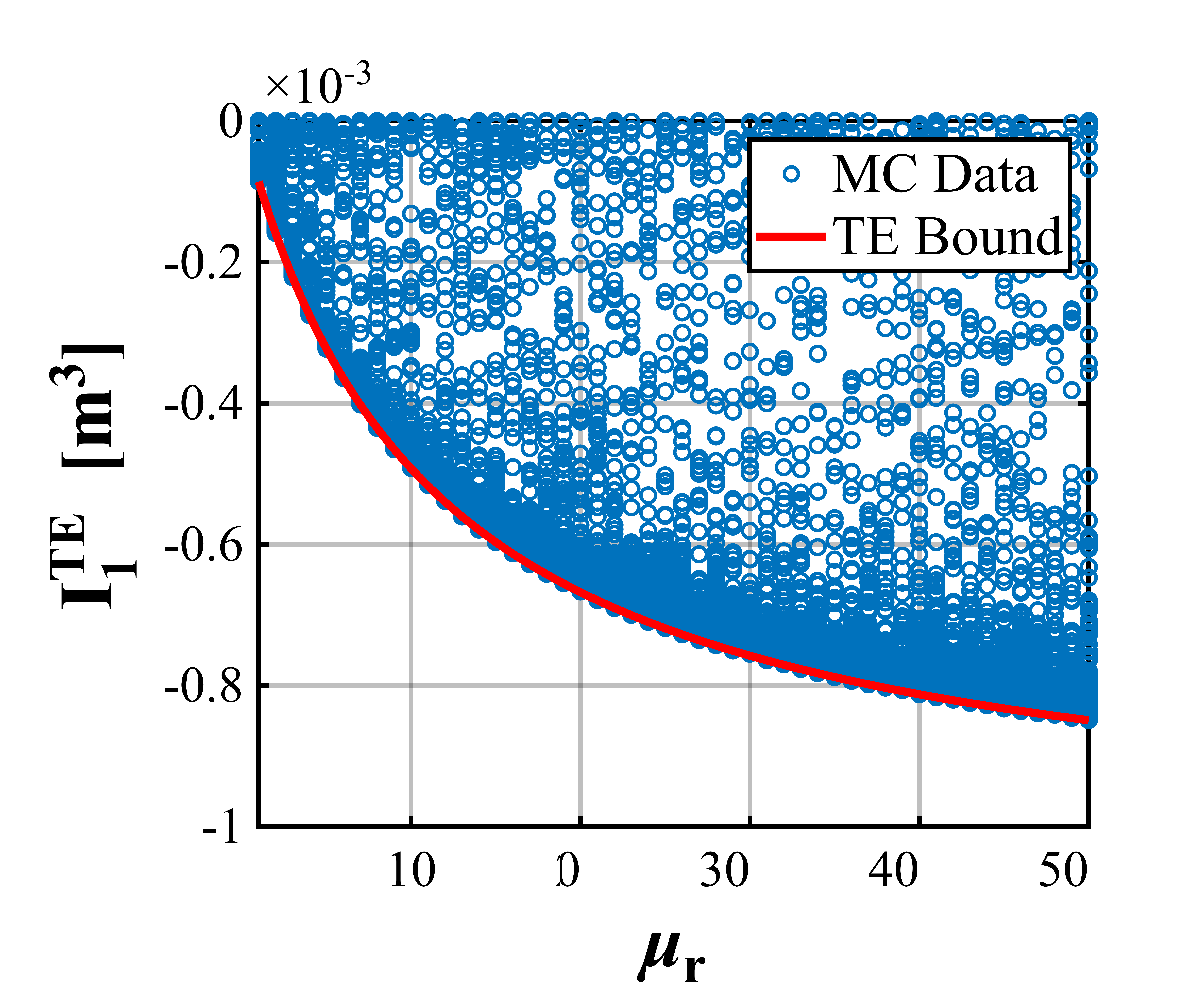}\\[-1mm]
    \hspace{3mm}(g)
\end{minipage}
\hfill
\begin{minipage}{0.31\textwidth}
    \centering
    \includegraphics[width=\linewidth]{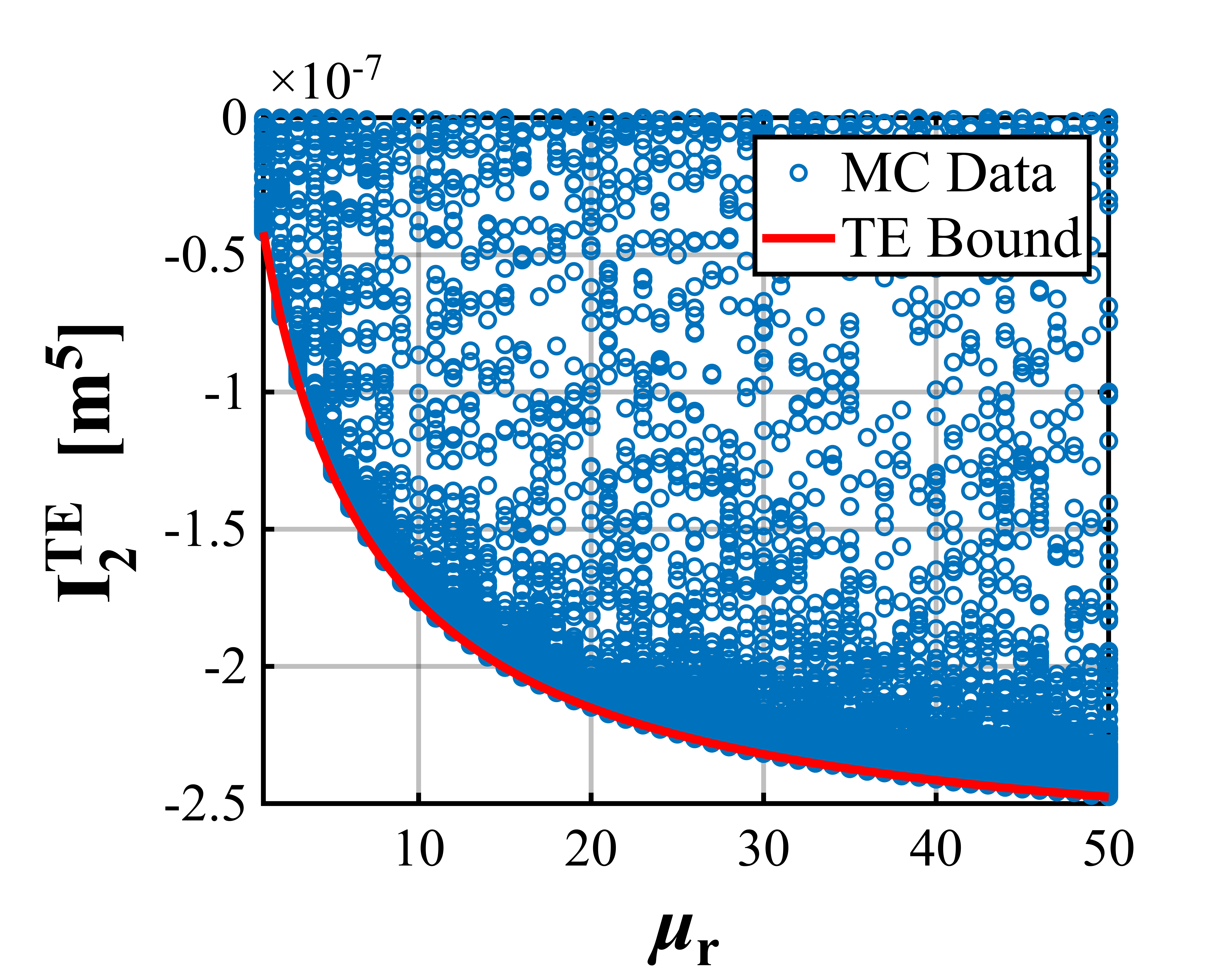}\\[-1mm]
    \hspace{3mm}(h)
\end{minipage}
\hfill
\begin{minipage}{0.31\textwidth}
    \centering
    \includegraphics[width=\linewidth]{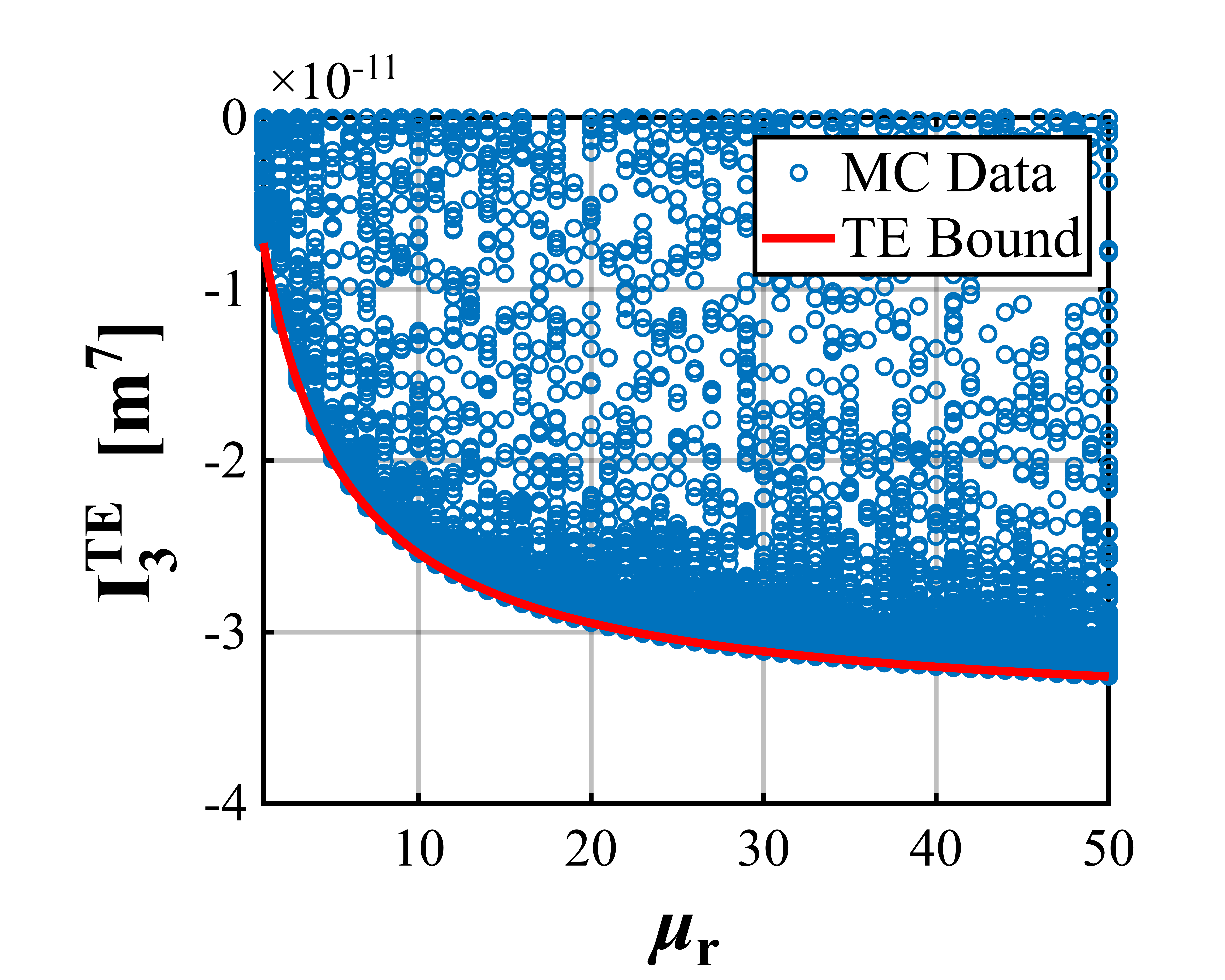}\\[-1mm]
    \hspace{3mm}(i)
\end{minipage}

\vspace{1.5mm}

\begin{minipage}{0.31\textwidth}
    \centering
    \includegraphics[width=\linewidth]{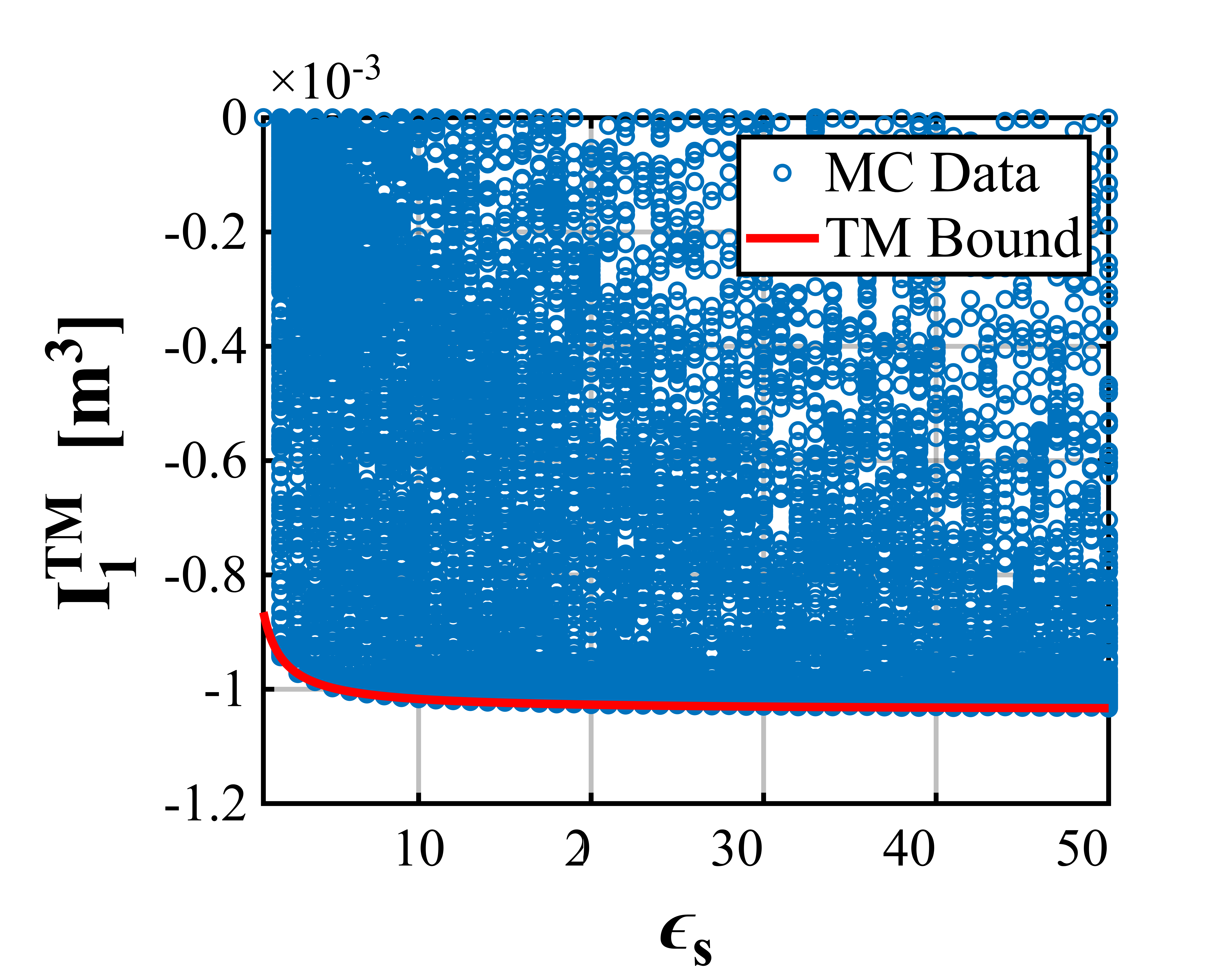}\\[-1mm]
    \hspace{3mm}(j)
\end{minipage}
\hfill
\begin{minipage}{0.31\textwidth}
    \centering
    \includegraphics[width=\linewidth]{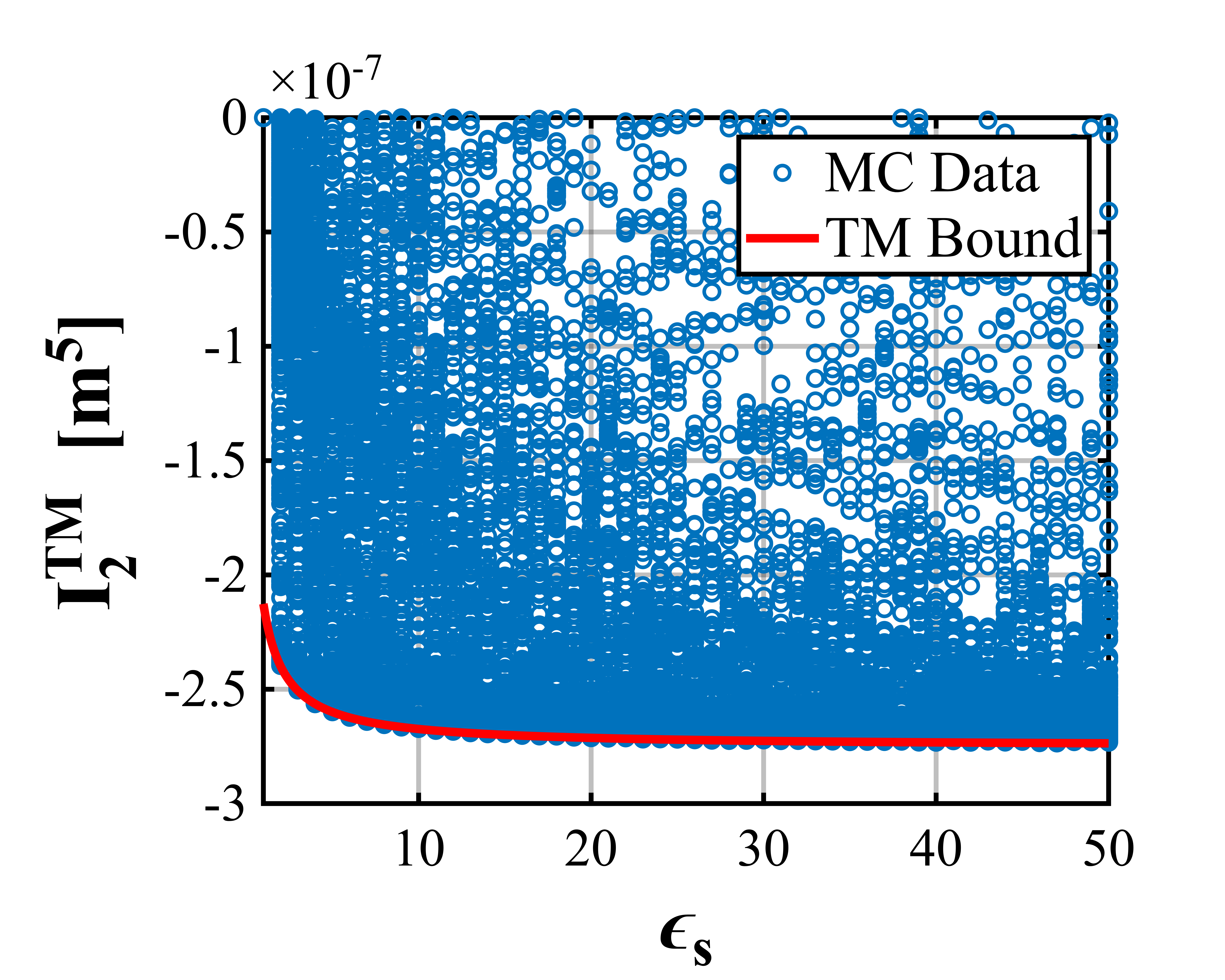}\\[-1mm]
    \hspace{3mm}(k)
\end{minipage}
\hfill
\begin{minipage}{0.31\textwidth}
    \centering
    \includegraphics[width=\linewidth]{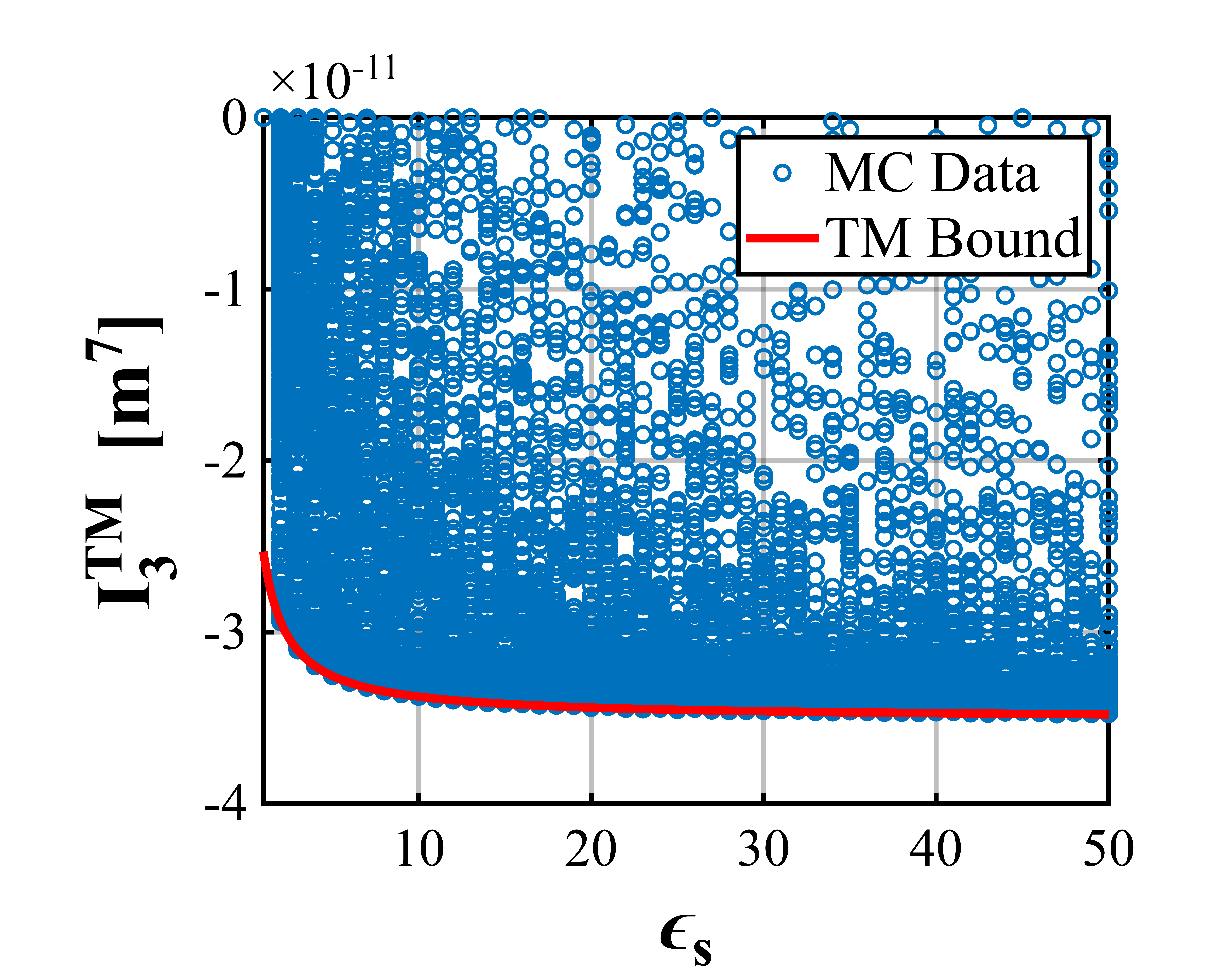}\\[-1mm]
    \hspace{3mm}(l)
\end{minipage}

\caption{Monte Carlo tests of the absorption bound for $n=1,2,3$.
(a)--(c) TE modes with swept constant $\mu_r$ and Drude-dispersive
permittivity; (d)--(f) TM modes with swept constant $\epsilon_r$ and
Drude-dispersive permeability; (g)--(i) TE modes using the common
resonant-Lorentz permittivity model with swept constant $\mu_r$; and
(j)--(l) TM modes using the same resonant-Lorentz permittivity model,
with the horizontal axis corresponding to the prescribed static
permittivity $\epsilon_s$. Markers denote the Monte Carlo realizations
and red curves denote the corresponding analytical bounds.}
\label{fig:mc_all}

\end{figure*}

The results are shown in
Figs.~\ref{fig:mc_all}(a-c) and~\ref{fig:mc_all}(d-f).
In each panel, the blue markers denote the Monte Carlo samples and the red curve
denotes the analytical bound $-\pi B_n^p/2$. The horizontal axis is
\(\mu_r\) for the TE modes and \(\epsilon_r\) for the TM modes. 

The preceding simulation used complementary dispersive models: $\epsilon_r$ was
dispersive for the TE modes, whereas $\mu_r$ was
dispersive for the TM modes. A second simulation used a
resonant-Lorentz permittivity model for both polarizations,
\begin{equation}
\epsilon_r(\omega)
=
\epsilon_\infty-
\frac{\omega_p^2}{\omega^2-\omega_0^2-j\Gamma_e\omega},
\qquad
\mu_r(\omega)=\mu_\infty,
\label{eq:common_material_model}
\end{equation}
where $\mu_r$ is real and frequency independent. The parameter ranges
were
\begin{equation*}
\epsilon_\infty,\mu_\infty\!\in\![1,10^6],
\mbox{ }
\omega_p,\Gamma_e\!\in\![1,10^{12}]\mbox{rad/s },
\omega_0\!\in\![1,10^{10}]\mbox{rad/s }
\label{eq:common_parameter_ranges}
\end{equation*}
with $a=1 ~\mathrm{cm}$, $b/a=1.1$.
For the TE modes, the material model in
Eq.~\eqref{eq:common_material_model} was used with the constant permeability
swept in $\mu_r=1...50$, while \(\epsilon_\infty\),
\(\omega_p\), \(\omega_0\), and \(\Gamma_e\) were independently randomized
over the ranges above. The results are shown in Fig.~\ref{fig:mc_all}(g-i).

For the TM modes, the swept parameter was the static relative
permittivity
\begin{equation}
\begin{aligned}
\epsilon_s
&=\epsilon_r(0)
=\epsilon_\infty+
\left(\frac{\omega_p}{\omega_0}\right)^2,
\\
\epsilon_s&=N,
\qquad N=1,2,\ldots,50.
\end{aligned}
\label{eq:common_static_permittivity}
\end{equation}
For each prescribed value \(N\), the parameters were generated as
\begin{equation}
\begin{aligned}
\epsilon_\infty&\in[1,N],
&\mu_\infty&\in[1,10^6],
\\
\omega_p,\Gamma_e&\in[1,10^{12}]~\mathrm{rad/s},
&\omega_0&=\frac{\omega_p}{\sqrt{N-\epsilon_\infty}}.
\end{aligned}
\label{eq:common_pinned_resonance}
\end{equation}
This construction guarantees \(\epsilon_r(0)=N\). The case $N=1$ is the nondispersive limiting realization $\epsilon_\infty=1$, $\omega_p=0$.
In contract with the TE case where $\omega_0$ was randomized, for the TM case, \(\omega_0\) is fixed by
Eq.~\eqref{eq:common_pinned_resonance}. The TM bound was evaluated using the prescribed static
permittivity \(\epsilon_s=N\) defined in
Eq.~\eqref{eq:common_static_permittivity}.
The Monte-Carlo results for this scenario are shown in
Fig.~
\ref{fig:mc_all}(j-l).

The numerical tests are not intended as a proof of
Conjecture~1.  Nevertheless, they provide a nontrivial test of the
proposed higher-order constraint.  In particular, the weighted
integral is sensitive both to narrow modal resonances and to the
long-wavelength region, where $|R_n^p|\rightarrow 1$.  Across the
large ensemble of passive dispersive realizations considered here,
including both Drude and resonant-Lorentz material models, no
counterexample to Eq.~(\ref{eq:bound_n}) was found.  Moreover, a substantial
fraction of the realizations approach the analytical boundary,
indicating that the numerical test probes the vicinity of the
proposed constraint rather than only a weakly absorbing regime.

\section{Geometrical Interpretation and Connection to Planar Bounds}
\label{sec:geometrical_interpretation}

The modal bounds derived above admit a useful geometrical
interpretation. In contrast to a planar PEC-backed absorber, for
which the relevant geometrical scale is primarily the coating
thickness, the spherical problem introduces an additional length
scale associated with curvature. This leads to a hierarchy of
bounds indexed by the spherical-harmonic order $n$. In particular,
the long-wavelength behavior
\(
R_n^p-1=O\left[(k_0b)^{2n+1}\right]
\)
results in a weighted reflection integral of the form in Eq.~(\ref{eq:bound_n}) rather than the unweighted wavelength integral appearing in the
usual planar Rozanov bound. Thus, curvature does not merely modify
the numerical constant in a planar thickness--bandwidth relation;
it separates the response into \emph{angular-momentum} channels, each
characterized by a different long-wavelength moment.

This geometrical dependence is particularly apparent through the
factor
\begin{equation}
\rho^{2n+1},\qquad \rho=\frac{a}{b},
\label{eq:rho_geometrical}
\end{equation}
which enters both $C_{\rm TE}$ and $C_{\rm TM}$ in Eq.~(\ref{eq:Cm}). 
In the quasistatic
limit, the two independent radial solutions of a spherical
multipole are proportional to $r^n$ and $r^{-(n+1)}$. Propagation
between the inner and outer interfaces therefore naturally
introduces the ratio $(a/b)^{2n+1}$. Accordingly,
$\rho^{2n+1}$ may be interpreted as a measure of the quasistatic
coupling of the $n$th spherical mode between the outer surface and
the inner PEC boundary.
This interpretation becomes especially transparent for a thin
coating. Let
\begin{equation}
d=b-a,\qquad d/b\ll1.
\end{equation}
Then
\begin{equation}
\left(\frac{a}{b}\right)^{2n+1}
=
\left(1-\frac{d}{b}\right)^{2n+1}
\simeq
\exp\left[-(2n+1)\frac{d}{b}\right].
\label{eq:spherical_exponential}
\end{equation}
On the other hand, a spherical harmonic of order $n$ possesses an
effective tangential wavenumber
\begin{equation}
k_t=\frac{\sqrt{n(n+1)}}{b}
\simeq\frac{n+1/2}{b},
\label{eq:effective_kt}
\end{equation}
where the last expression represents the large-$n$ limit.
Consequently,
\begin{equation}
\left(\frac{a}{b}\right)^{2n+1}
\simeq e^{-2k_t d}.
\label{eq:planar_correspondence}
\end{equation}
Equation~(\ref{eq:planar_correspondence}) provides a direct
geometrical connection between spherical and planar problems.
Angular momentum on the sphere plays the role of transverse
momentum in a planar structure, while the power-law coupling
$\rho^{2n+1}$ approaches the familiar exponential dependence
associated with radial penetration of a field having transverse
wavenumber $k_t$.

More formally, consider the locally planar limit
\begin{equation}
b\rightarrow\infty,\qquad n\rightarrow\infty,
\qquad \frac{n+1/2}{b}\rightarrow k_t,
\end{equation}
while the physical coating thickness $d=b-a$ remains fixed. In this
limit,
\begin{equation}
\lim_{\substack{b,n\rightarrow\infty\\(n+1/2)/b=k_t}}
\left(1-\frac{d}{b}\right)^{2n+1}
=e^{-2k_td}.
\label{eq:local_planar_limit}
\end{equation}
Thus, the discrete spherical index $n$ continuously approaches the
transverse-wavenumber parameter of a locally planar problem. This
observation suggests viewing the present bounds as the
angular-momentum-resolved counterpart of planar broadband
absorption constraints. Establishing an exact reduction to a
particular planar Rozanov bound additionally requires matching the
normalization and excitation of the spherical and planar channels
and is not implied by (\ref{eq:local_planar_limit}) alone.

Further insight follows directly from the thin-coating expansion.
Using
\begin{equation}
1-\rho^{2n+1}\simeq(2n+1)\frac{d}{b},
\end{equation}
the TE coefficient becomes
\begin{equation}
C_{\rm TE}
\simeq
\frac{b}{\mu_r d},
\label{eq:CTE}
\end{equation}
and hence, to leading order in $d/b$,
\begin{equation}
B_n^{\rm TE}
\simeq
\frac{2(2\pi)^{2n+1}}
{\left[(2n-1)!!\right]^2}
\,\mu_r b^{2n}d.
\label{eq:thin_TE_bound}
\end{equation}
The linear dependence on $d$ is reminiscent of the thickness
dependence of planar Rozanov-type bounds, whereas the factor
$b^{2n}$ reflects the multipolar geometry of the finite object.
For the dipolar channel, $n=1$, this scaling reduces to
\begin{equation}
B_1^{\rm TE}\propto \mu_r b^2d.
\end{equation}
Since $4\pi b^2d$ is the volume of a thin spherical shell to leading
order, the coefficient governing the dipolar TE absorption bound can
therefore be interpreted, up to a numerical factor, as scaling with the
magnetostatic material response times the available coating volume.

The TM thin-shell limit is qualitatively different. In this case,
\begin{equation}
C_{\rm TM}
\simeq
\frac{n(n+1)}{\epsilon_r}\frac{d}{b},
\end{equation}
so that
{
\begin{equation}
B_n^{\rm TM}
\rightarrow
A_n\frac{2n+1}{n},
\qquad d/b\rightarrow0.
\label{eq:thin_TM_bound}
\end{equation}
}%
Thus, unlike the TE bound, the right-hand side does not vanish with
the coating thickness. This should not be interpreted as finite
absorption by a vanishing coating: for a bare lossless PEC sphere,
$|R_n^p|=1$ and the absorption integral itself vanishes. Rather,
(\ref{eq:thin_TM_bound}) indicates that the present causal bound is
not asymptotically tight with respect to coating thickness for the
TM channel. Additional constraints associated with the available
lossy material may therefore be required to obtain a tight
ultrathin-layer TM bound. 
The qualitative difference between the TE and TM thin-coating limits
is physically consistent with the distinct quasistatic character of
the two polarizations. In the TE case, the electrically small PEC
sphere exhibits an inductive modal response, whereas in the TM case
the corresponding response is capacitive. A closely related
distinction was identified in \cite{firestein2023} for a partially
transparent Dallenbach absorber. There, the addition of a capacitive
impedance sheet was shown to eliminate the corresponding finite
absorption bound, whereas an inductive sheet preserves the bound,
although in a relaxed form. This analogy further suggests that the
different thin-coating behavior of the TE and TM bounds is rooted in
their fundamentally different quasistatic loading.
The role of the PEC core can also be identified by considering
$a/b\rightarrow0$. In this limit,
{
\begin{equation}
C_{\rm TM}\rightarrow\frac{n+1}{\epsilon_r},
\qquad
C_{\rm TE}\rightarrow\frac{n+1}{\mu_r},
\end{equation}
}%
and consequently
{
\begin{equation}
B_n^p\rightarrow
A_n\frac{2n+1}{n+(n+1)/\chi_p},
\qquad
\chi_{\rm TM}=\epsilon_r,\quad
\chi_{\rm TE}=\mu_r.
\label{eq:small_core}
\end{equation}
}%
The TE and TM bounds therefore acquire a {similar} form in which the
electric and magnetic static responses enter symmetrically.
Moreover, in the large-static-response limit
$\chi_p\rightarrow\infty$,
{
\begin{equation}
B_n^p\rightarrow
A_n\frac{2n+1}{n}.
\label{eq:geometrical_saturation}
\end{equation}
}%
This saturation is not restricted to the small-core limit: for any $a/b$, $C_p\to0$ as $\chi_p\to\infty$, and Eq.~\eqref{eq:geometrical_saturation} follows.
The bound then becomes independent of further increases in the
constitutive parameter and is controlled entirely by the geometry
of the spherical channel.

The saturation in Eq.~\eqref{eq:geometrical_saturation} has no counterpart in
the planar problem, and comparing the two limits directly renders the role of
curvature quantitative. Consider the same broadband specification
$|R|\le\rho_0$ over $\lambda\in[\lambda_1,\lambda_2]$ imposed, on the one hand,
on a planar PEC-backed layer of thickness $d$ and static permeability $\mu_r$
and, on the other, on the TE$_n$ channel of a coated sphere with the same $d$
and $\mu_r$. Each geometry therefore defines a \emph{maximum permitted logarithmic reflection reduction} --- the largest
value of $\ln(1/\rho_0)$ allowed by its bound. Writing the corollary
\eqref{eq:rozanov_corollary} as $\ln(1/\rho_0)\le E_n^{p}$, with
\begin{equation}
E_n^{p}
=\frac{\pi(2n+1)\,B_n^{p}}
       {2\bigl(\lambda_2^{2n+1}-\lambda_1^{2n+1}\bigr)},
\label{eq:En_exponent}
\end{equation}
and the classical planar limit \cite{rozanov} in the same form,
$\ln(1/\rho_0)\le E_R=2\pi^2\mu_r d/(\lambda_2-\lambda_1)$, the comparison
reduces to the single dimensionless ratio
\begin{equation}
\Gamma_n=\frac{E_n^{\rm TE}}{E_R}.
\label{eq:Gamma_def}
\end{equation}
The meaning of $\Gamma_n$ is fixed once and for all by the sign of
$\Gamma_n-1$: if $\Gamma_n>1$ the spherical channel permits a larger
$\ln(1/\rho_0)$, so for the same $d$ and $\mu_r$ it \emph{permits a lower
in-band reflection $\rho_0$} than the planar layer; if $\Gamma_n<1$ the
planar layer permits the larger $\ln(1/\rho_0)$, and the spherical channel
is the more demanding of the two. In the figures below this is simply whether
a TE curve lies below or above the planar curve.

\begin{figure*}[t]
  \centering
  \includegraphics[width=\linewidth]{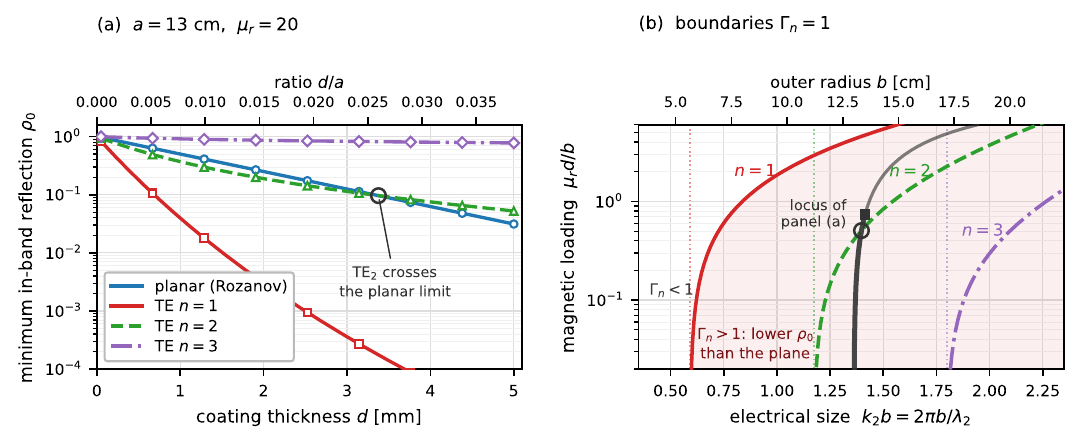}
  \caption{Curvature versus the planar thickness--bandwidth limit, band
0.5--10~GHz ($\lambda_2 = 60$~cm), $b = a + d$.
[TE channels only; see the remark following Eq.~\eqref{eq:gamma_thin}.]
(a)~Minimum in-band reflection $\rho_0$ permitted by the planar limit and by the
TE$_n$ channels, for $a = 13$~cm and $\mu_r = 20$; a curve lying below the planar
curve marks a channel that permits a lower $\rho_0$, i.e.\ $\Gamma_n > 1$.
TE$_1$ permits a far lower $\rho_0$ throughout; TE$_2$ starts below the planar
curve and crosses it (circle, $d \simeq 3.4$~mm) as its geometric ceiling
\eqref{eq:geometrical_saturation} is approached; TE$_3$, for which
$b<b_{\rm th}^{(3)}$,
is the more demanding channel at every thickness.
(b)~The boundaries $\Gamma_n = 1$ of Eq.~\eqref{eq:gamma_thin} in the
electrical-size/loading plane. Below and to the right of a boundary that channel
permits the lower $\rho_0$ ($\Gamma_n > 1$, shaded for $n = 1$); in the unshaded
region every channel is the more demanding one.
Dotted verticals mark $b_{\rm th}^{(n)}$ of Eq.~\eqref{eq:ath}, where each
boundary meets the zero-loading axis; at finite loading the boundary itself lies
to the right of its vertical.
The heavy black segment is the locus traced by panel~(a) as $d$ runs from $0.05$
to $5$~mm at fixed $a = 13$~cm and $\mu_r = 20$, ending at the square; where it
crosses the $n = 2$ boundary (circle) is the crossing marked in panel~(a). The
thin line continues the same sweep in $d$, at the same $a$ and $\mu_r$, up to the
top of the frame, which the loading reaches at $d = 55.7$~mm ($b = 18.6$~cm). The
locus leans to the right because $b = a + d$ grows with the coating thickness;
along all of it the thin-coating form Eq.~\eqref{eq:gamma_thin} stays within
$2.1\,\%$
of the ratio formed from the exact coefficient $B_n^p$ of \eqref{eq:Bn}.}
\label{fig:curvature_vs_planar}
\end{figure*}

{For a thin coating substituting Eq.~\eqref{eq:CTE} in \eqref{eq:Gamma_def} gives
\begin{equation}
  \Gamma_n = \frac{\bigl(b/b_{\rm th}^{(n)}\bigr)^{2n}}{1 + n\,\mu_r d/b}\,,
  \label{eq:gamma_thin}
\end{equation}
where
\begin{equation}
  k_2 b_{\rm th}^{(n)} = \frac{\bigl[(2n-1)!!\bigr]^{1/n}}{\bigl[(2n+1)W_n\bigr]^{1/2n}}\,,
  \qquad
  W_n = \frac{1-r}{1-r^{2n+1}}\,,
  \label{eq:ath}
\end{equation}
with $k_2 = 2\pi/\lambda_2$ and $r = \lambda_1/\lambda_2$. Here $b_{\rm th}^{(n)}$ is the sphere radius at which the Rozanov and the $n$th
spherical-channel limits coincide for a lightly loaded coating, i.e.\
$\Gamma_n = 1$ when $b = b_{\rm th}^{(n)}$ and $\mu_r d \ll b$. It follows form Eq.~\eqref{eq:ath} that $b_{\rm th}^{(n)}$ is fixed by the design band and the channel index alone and not by the coating. For the band 0.5--10~GHz,
$b_{\rm th}^{(1)} = 5.7$~cm, $b_{\rm th}^{(2)} = 11.2$~cm and
$b_{\rm th}^{(3)} = 17.2$~cm.

Two consequences follow, both visible in Fig.~\ref{fig:curvature_vs_planar}. First, the material enters \eqref{eq:gamma_thin} only through the factor $(1+n\mu_rd/b)^{-1}$, and the two limits respond to loading in opposite ways. The spherical quantity $E_n^{\rm TE}$ saturates: it approaches the purely geometric ceiling implied by \eqref{eq:geometrical_saturation}, reaching half of its saturation value already at $\mu_rd=b/n$. The planar quantity $E_R$ does not saturate: it grows linearly in $\mu_rd$ without bound. Increasing magnetic loading therefore produces progressively smaller increases in $E_n^{\rm TE}$ than in $E_R$, and $\Gamma_n$ decreases monotonically as the coating is loaded. Second, a
channel permits a lower $\rho_0$ than the planar layer only if the sphere is
electrically large enough and the coating lightly enough loaded,
\[
  b > b_{\rm th}^{(n)}
  \qquad\text{and}\qquad
  \mu_r d < \frac{b}{n}\left[\left(\frac{b}{b_{\rm th}^{(n)}}\right)^{2n} - 1\right].
\]
The two conditions act together, so $b_{\rm th}^{(n)}$ should be read as the
lightly loaded limit and not as the crossing itself: at finite loading the
crossing sits at
$b^{\ast} = b_{\rm th}^{(n)}\bigl(1 + n\mu_r d/b^{\ast}\bigr)^{1/2n}$, which for
the parameters of Fig.~\ref{fig:curvature_vs_planar}(a) is $1.37\,b_{\rm th}^{(1)}$, $1.19\,b_{\rm th}^{(2)}$
and $1.13\,b_{\rm th}^{(3)}$.

It is instructive to explore the significance of $b_{\rm th}^{(n)}$. To that end, noting that both factors in Eq.~\eqref{eq:gamma_thin} vary smoothly with the sphere radius, and $\Gamma_n$
grows steadily with it, as discussed, $b_{\rm th}^{(n)}$ is where a ratio passes smoothly through unity. {The numerator captures the curvature-dependent geometric factor. On a small sphere a channel of order $n$ couples only weakly to the coating, this is the $(k_0b)^{2n+1}$ behavior of $R_n^p-1$, which gives the $b^{2n}$ in Eq.~\eqref{eq:thin_TE_bound} that the planar limit does not contain. It is that factor, written as $(b/b_{\rm th}^{(n)})^{2n}$ in Eq.~ \eqref{eq:gamma_thin}, that grows with the sphere.}
The denominator captures the saturation with loading: by Eq.~\eqref{eq:geometrical_saturation} the spherical limit
saturates as material is added, whereas $E_R$ does not. A channel permits the
lower $\rho_0$ once the geometric factor in the numerator is sufficiently large
that $\Gamma_n\!>\!1$, which for a lightly loaded coating happens at $b = b_{\rm th}^{(n)}$. Below that radius
the channel is coupled too weakly to the coating for $\Gamma_n\!>\!1$, and the spherical channel is by a wide margin the more demanding of the
two; above it, $\Gamma_n\!>\!1$ becomes possible for sufficiently light loading. Since $b_{\rm th}^{(n)}$ grows with
$n$, a sphere of increasing size admits the successive channels into the favorable
regime one at a time.}

{Following the discussion, it is noted that only the TE channels admit a meaningful comparison with $E_R$. This is because the TM spherical limit does not vanish with the coating thickness: by Eq.~\eqref{eq:thin_TM_bound} it tends to the ceiling of Eq.~\eqref{eq:geometrical_saturation}, for any $\epsilon_r$ and any thickness. Equation \eqref{eq:Gamma_def}, formulated for TM, then ceases to be informative: its numerator tends to a constant while its denominator vanishes with $d$, so $\Gamma_n^{\rm TM}$ grows without bound. This is the non-tightness noted after Eq.~\eqref{eq:thin_TM_bound}. And there is no $\epsilon_r$ in $E_R$ to substitute: the planar limit contains the static permeability alone. Hence the similarity in Eq.~\eqref{eq:small_core} is between the two spherical bounds and does not extend to $E_R$.}

Taken together, these limits show that curvature plays a more
fundamental role than providing a perturbative correction to the
planar thickness--bandwidth limit. It introduces a discrete
transverse-momentum spectrum, with $n/b$ serving as the spherical
counterpart of planar transverse wavenumber, and controls the
coupling to the inner PEC surface through
$(a/b)^{2n+1}$. In the locally planar limit this coupling approaches
$e^{-2k_td}$, providing a direct connection between angular
momentum on a curved surface and transverse momentum in planar
absorber theory.

\section{Absorption bound for a localized external source}
\label{sec:finite_source}

The modal bounds derived above can also be used to constrain the
broadband absorption produced by a spatially localized source outside
the coated sphere. To demonstrate this connection, we consider a
configuration that isolates a particularly simple subset of the
spherical channels, although the derivation and equations are general.

\begin{figure*}[t]
    \centering
    \includegraphics[width=0.98\textwidth]
    {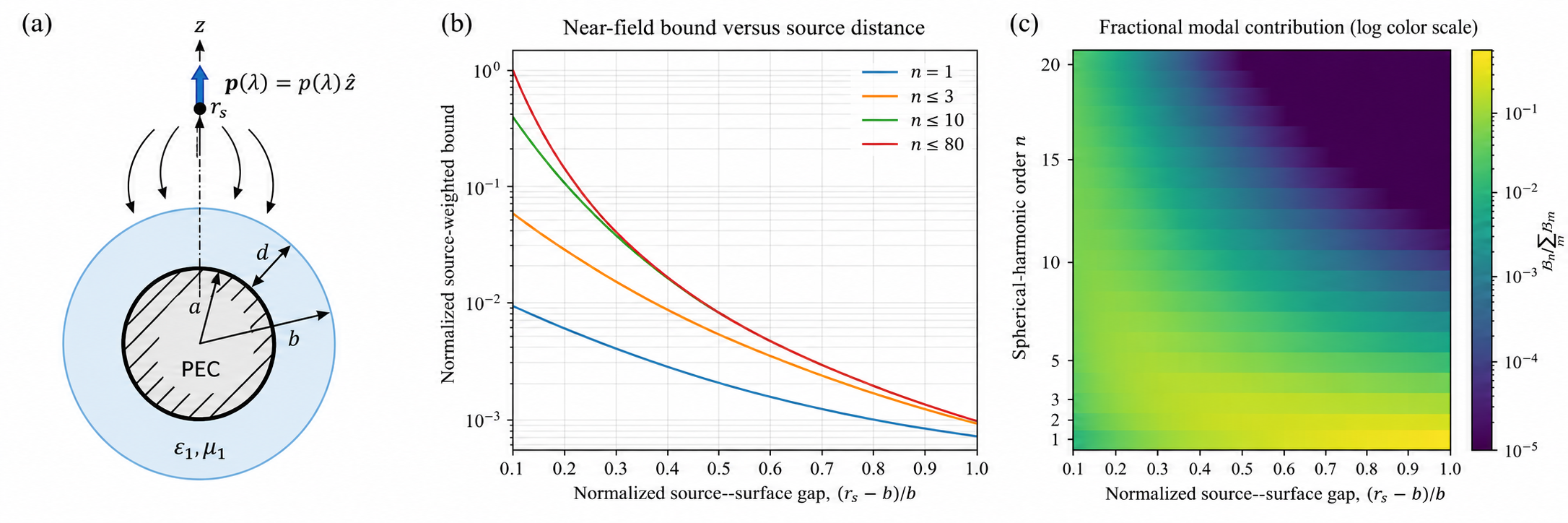}
    \caption{Source-weighted broadband absorption bound for the radial
    electric dipole of
    Eqs.~\eqref{eq:source_position}--\eqref{eq:source_dipole}.
    (a) The geometry is $a=1~\mathrm{cm}$, $b/a=1.1$, and
    $\epsilon_s=10$, and the source band is $1$--$3~\mathrm{GHz}$.
    The spherical-wave source weights follow
    Eq.~\eqref{eq:source_modal_weight}, with the common spectral
    factor $q_0(\lambda)$ specified by
    Eq.~\eqref{eq:flat_source_spectrum}. Since the sphere lies on
    the radiation null of the radial dipole, only the near-field
    range $0.1\leq(r_s-b)/b\leq1$ is considered.
    (b) Cumulative bound $\sum_{n=1}^{N}\mathcal B_n$, where
    $\mathcal B_n$ is defined in Eq.~\eqref{eq:source_Bn}, for the
    indicated truncation orders. The curves are normalized by an
    $n\leq120$ reference at $(r_s-b)/b=0.1$; the $n\leq80$ result
    differs from this reference by less than $0.003\%$.
    (c) Fractional contribution
    $\mathcal B_n/\sum_m\mathcal B_m$ of the first 20 spherical
    orders, shown on a logarithmic color scale. At small separation
    the localized near field occupies a broad range of angular
    momenta; with increasing separation the high-$n$ contributions
    are suppressed and the bound becomes concentrated in the lowest
    spherical channels, consistently with the effective
    transverse-wavenumber relation in Eq.~(\ref{eq:effective_kt}).}
    \label{fig:finite_source_bound}
\end{figure*}

Consider an electric dipole located on the positive $z$ axis,
\begin{equation}
    \mathbf r_s=r_s\hat{\mathbf z},
    \qquad r_s>b,
    \label{eq:source_position}
\end{equation}
and oriented radially (See illustration in Fig.~\ref{fig:finite_source_bound}(a)),
\begin{equation}
    \mathbf p(\lambda)=p(\lambda)\hat{\mathbf z}.
    \label{eq:source_dipole}
\end{equation}
The radial orientation is useful because, by axial symmetry, this
source excites only $m=0$ TM spherical waves. We emphasize that the
configuration is used here as a \emph{near-field} excitation example.
Indeed, the $1/r$ radiative field of a $z$-directed electric dipole
vanishes on its axis, whereas its induction and quasistatic fields,
proportional to $1/r^2$ and $1/r^3$, respectively, remain nonzero
\cite{balanis, harrington}. We therefore restrict the numerical example below to
source positions for which the sphere remains in the near-field region of the dipole.

For completeness, we briefly establish the spherical-wave expansion
used below. We employ normalized complex vector spherical harmonics
\cite{harrington,Stratton,Tai},
\begin{equation}
    \mathbf X_{nm}(\hat{\mathbf r})
    =
    \frac{\mathbf L Y_{nm}(\hat{\mathbf r})}
         {\sqrt{n(n+1)}},
    \qquad
    \int_{\Omega}|Y_{nm}|^2\,d\Omega=1,
    \label{eq:VSH_definition}
\end{equation}
where $\mathbf L=-j\mathbf r\times\nabla$. The corresponding vector
spherical waves are defined as
\begin{equation}
\begin{aligned}
    \mathbf M_{nm}^{(z)}(k\mathbf r)
    &=
    z_n(kr)\mathbf X_{nm}(\hat{\mathbf r}),
    \\
    \mathbf N_{nm}^{(z)}(k\mathbf r)
    &=
    \frac{1}{k}\nabla\times
    \mathbf M_{nm}^{(z)}(k\mathbf r).
\end{aligned}
\label{eq:VSWF_definition}
\end{equation}
where $z_n=j_n$, $h_n^{(1)}$, or $h_n^{(2)}$. With the
$e^{j\omega t}$ convention adopted throughout this work,
$h_n^{(1)}$ and $h_n^{(2)}$ represent inward- and
outward-propagating spherical waves, respectively.

The free-space electric dyadic Green function admits an eigenfunction
expansion in these vector spherical waves
\cite{Stratton,Tai}. For $r<r_s$, the radial dependence associated
with the observation coordinate is regular, whereas that associated
with the source coordinate is outgoing. Apart from a factor that is
common to all $(n,m)$ channels, the electric field of an electric
dipole in the domain $r<r_s$ can thus be written in the form
\begin{equation}
\begin{aligned}
\mathbf{E}^{\mathrm{src}}(\mathbf r,\lambda)\propto
\sum_{n=1}^{\infty}&\sum_{m=-n}^{n}
\Bigl[
\mathbf{M}_{nm}^{\mathrm{reg}}(k_0\mathbf r)\,
\mathbf{M}_{nm}^{\mathrm{out}*}(k_0\mathbf r_s)\cdot\mathbf p \\
&+\mathbf{N}_{nm}^{\mathrm{reg}}(k_0\mathbf r)\,
\mathbf{N}_{nm}^{\mathrm{out}*}(k_0\mathbf r_s)\cdot\mathbf p
\Bigr].
\end{aligned}
\label{eq:dyadic_green_expansion}
\end{equation}
The common prefactor in Eq.~\eqref{eq:dyadic_green_expansion} is not
needed below because it is collected into the common source spectral
weight introduced explicitly in Eq.~\eqref{eq:source_modal_weight}.

For the radial dipole of Eqs.~\eqref{eq:source_position} and
\eqref{eq:source_dipole}, the source vector is parallel to
$\hat{\mathbf r}_s$. Since $\mathbf M_{nm}$ has no radial component,
the first term of Eq.~\eqref{eq:dyadic_green_expansion} vanishes.
Moreover, on the positive $z$ axis,
\begin{equation}
    Y_{nm}(0,\phi)=0,\qquad m\neq0,
\end{equation}
while
\begin{equation}
    Y_{n0}(0,\phi)=\sqrt{\frac{2n+1}{4\pi}}.
\end{equation}
Consequently, only the TM, $m=0$ channels remain, and in the
source-free region $b<r<r_s$ the field has the expansion
\begin{equation}
    \mathbf E^{\rm src}(\mathbf r,\lambda)
    =
    \sum_{n=1}^{\infty}
    c_n(\lambda;r_s)
    \mathbf N_{n0}^{\rm reg}(k_0\mathbf r),
    \qquad b<r<r_s .
    \label{eq:source_vsw_expansion}
\end{equation}

The radial component of the TM vector spherical wave defined in
Eq.~\eqref{eq:VSWF_definition} is
\begin{equation}
    \hat{\mathbf r}\cdot
    \mathbf N_{nm}^{(z)}(k\mathbf r)
    =
    \sqrt{n(n+1)}
    \frac{z_n(kr)}{kr}
    Y_{nm}(\theta,\phi),
    \label{eq:N_radial_component}
\end{equation}
up to an overall phase that depends on the convention used for
$\mathbf X_{nm}$ and is immaterial here. Evaluating
Eq.~\eqref{eq:N_radial_component} at the source position therefore
gives
\begin{equation}
    \left|
    \hat{\mathbf z}\cdot
    \mathbf N_{n0}^{\rm out}(k_0r_s\hat{\mathbf z})
    \right|^2
    =
    \frac{n(n+1)(2n+1)}{4\pi}
    \left|
    \frac{h_n^{(2)}(k_0r_s)}
         {k_0r_s}
    \right|^2 .
    \label{eq:dipole_vsw_projection}
\end{equation}
Equations~\eqref{eq:dyadic_green_expansion}--%
\eqref{eq:dipole_vsw_projection} give directly the dependence of the
source coupling on both $n$ and $r_s$.

A regular wave contains equal inward- and outward-propagating
components,
\begin{equation}
    \mathbf N_{n0}^{\rm reg}
    =
    \frac{1}{2}
    \left(
        \mathbf N_{n0}^{\rm in}
        +
        \mathbf N_{n0}^{\rm out}
    \right),
    \label{eq:regular_decomposition}
\end{equation}
since $j_n=(h_n^{(1)}+h_n^{(2)})/2$. Hence the inward component
of the $n$th source-excited channel has amplitude
\begin{equation}
    a_n(\lambda;r_s)
    =
    \frac{1}{2}c_n(\lambda;r_s).
    \label{eq:source_inward_coeff}
\end{equation}

We denote by $q_n(\lambda;r_s)\geq0$ the inward modal energy
spectral density per unit wavelength. The factors that are common to
all spherical orders are collected into a single source spectral
function $q_0(\lambda)$. Using
Eq.~\eqref{eq:dipole_vsw_projection}, we therefore define
\begin{equation}
    q_n(\lambda;r_s)
    =
    q_0(\lambda)
    (2n+1)n(n+1)
    \left|
        \frac{h_n^{(2)}(k_0r_s)}
             {k_0r_s}
    \right|^2 .
    \label{eq:source_modal_weight}
\end{equation}
The factor $1/(4\pi)$ in
Eq.~\eqref{eq:dipole_vsw_projection}, together with the common
Green-function and source-amplitude factors, has been absorbed into
$q_0(\lambda)$. Thus Eq.~\eqref{eq:source_modal_weight} contains
the complete $n$- and $r_s$-dependence of the radial-dipole
excitation; $q_0(\lambda)$ specifies only the common spectral
weighting.

For the numerical example below, rather than prescribing a particular
frequency dependence of the physical dipole moment, we specify this
common spectral weighting directly:
\begin{equation}
    q_0(\lambda)
    =
    \begin{cases}
        q_0, & \lambda_1\leq\lambda\leq\lambda_2,\\
        0,   & \text{otherwise},
    \end{cases}
    \qquad q_0>0.
    \label{eq:flat_source_spectrum}
\end{equation}
Thus the common source weighting is constant \emph{per unit
wavelength} over the prescribed band, and no additional
frequency-dependent factor is implicit in the calculation. Since the
results plotted below are normalized, the constant $q_0$ cancels and
can be set to unity.

Since the coated sphere is spherically symmetric, the different
$(n,m)$ channels do not mix. The outgoing amplitude produced by an
inward TM amplitude $a_n$ is $R_n^{\rm TM}a_n$, so the fraction of
the inward modal energy absorbed in the $n$th channel is
\begin{equation}
    A_n(\lambda)
    =
    1-\left|R_n^{\rm TM}(\lambda)\right|^2 .
    \label{eq:source_modal_absorptance}
\end{equation}
The broadband energy absorbed from the finite source over
$\mathcal B=[\lambda_1,\lambda_2]$ is consequently
\begin{equation}
    \mathcal E_{\rm abs}
    =
    \sum_{n=1}^{\infty}
    \int_{\mathcal B}
    q_n(\lambda;r_s)
    \left[
        1-\left|R_n^{\rm TM}(\lambda)\right|^2
    \right]d\lambda .
    \label{eq:source_abs_energy}
\end{equation}

To connect Eq.~\eqref{eq:source_abs_energy} to the modal bound of
Sec.~III, define
\begin{equation}
    L_n(\lambda)
    =
    \ln\frac{1}{|R_n^{\rm TM}(\lambda)|}
    \geq0 .
    \label{eq:source_Ln}
\end{equation}
Equation~(\ref{eq:bound_n}) is equivalently
\begin{equation}
    \int_0^\infty
    \lambda^{2n}L_n(\lambda)\,d\lambda
    \leq
    \frac{\pi}{2}B_n^{\rm TM}.
    \label{eq:source_log_bound}
\end{equation}
Furthermore,
\begin{equation}
    1-\left|R_n^{\rm TM}\right|^2
    =
    1-e^{-2L_n}
    \leq
    2L_n .
    \label{eq:source_exp_ineq}
\end{equation}
Hence,
\begin{align}
    \mathcal E_{{\rm abs},n}
    &=
    \int_{\mathcal B}
    q_n(\lambda;r_s)
    \left(
        1-e^{-2L_n(\lambda)}
    \right)d\lambda
    \nonumber\\
    &\leq
    2
    \max_{\lambda\in\mathcal B}
    \left\{
        \frac{q_n(\lambda;r_s)}{\lambda^{2n}}
    \right\}
    \int_{\mathcal B}
    \lambda^{2n}L_n(\lambda)\,d\lambda
    \nonumber\\
    &\leq
    \pi B_n^{\rm TM}
    \max_{\lambda\in\mathcal B}
    \left\{
        \frac{q_n(\lambda;r_s)}{\lambda^{2n}}
    \right\}.
    \label{eq:source_channel_bound}
\end{align}
The resulting finite-source bound is therefore
\begin{equation}
{
    \mathcal E_{\rm abs}
    \leq
    \pi
    \sum_{n=1}^{\infty}
    B_n^{\rm TM}
    \max_{\lambda\in\mathcal B}
    \left\{
        \frac{q_n(\lambda;r_s)}{\lambda^{2n}}
    \right\}.
    }
    \label{eq:source_total_bound}
\end{equation}
As in Eq.~(\ref{eq:bound_n}), the $n=1$ contribution follows from
Theorem~1, while the contributions with $n\geq2$ rely on
Conjecture~1.

For convenience, define the contribution of the $n$th channel to the
right-hand side of Eq.~\eqref{eq:source_total_bound} as
\begin{equation}
    \mathcal B_n(r_s)
    \equiv
    \pi B_n^{\rm TM}
    \max_{\lambda\in\mathcal B}
    \left\{
        \frac{q_n(\lambda;r_s)}{\lambda^{2n}}
    \right\},
    \label{eq:source_Bn}
\end{equation}
so that
\begin{equation}
    \mathcal E_{\rm abs}
    \leq
    \sum_{n=1}^{\infty}\mathcal B_n(r_s).
    \label{eq:source_sum_bound}
\end{equation}

Equation~\eqref{eq:source_total_bound} shows how the individual modal
absorption bounds combine for a localized near-field source.
The source position enters solely through the spatial modal weights in
Eq.~\eqref{eq:source_modal_weight}. As the dipole approaches the
surface, its field varies more rapidly along the sphere and significant
weight is transferred to progressively larger angular momenta.
Conversely, increasing the source--surface separation suppresses the
higher-order components. In view of Eq.~(\ref{eq:effective_kt}), this behavior is the spherical-wave counterpart of the large
transverse-wavenumber content of a localized near field.

Figure~\ref{fig:finite_source_bound} illustrates the result for
$a=1~\mathrm{cm}$, $b/a=1.1$, $\epsilon_s=10$, and the frequency
band $1$--$3~\mathrm{GHz}$, corresponding to
$\lambda_1=c/(3~\mathrm{GHz})$ and
$\lambda_2=c/(1~\mathrm{GHz})$. Because the sphere lies on the
radiation null of the radial dipole, the example is deliberately
restricted to
\begin{equation}
    0.1
    \leq
    g\equiv\frac{r_s-b}{b}
    \leq
    1,
    \label{eq:source_gap_range}
\end{equation}
which for the present geometry corresponds to physical
source--surface gaps from $1.1~\mathrm{mm}$ to $11~\mathrm{mm}$.

Figure~\ref{fig:finite_source_bound}(b) shows the cumulative bound
$\sum_{n=1}^{N}\mathcal B_n$ for several truncation orders. The
curves are normalized by a numerical reference obtained by summing
through $n=120$ at $g=0.1$. At this smallest separation, the first
ten modes provide approximately $37\%$ of the total bound, whereas
orders through $n=40$ provide approximately $98.5\%$. The
$n\leq80$ result differs from the $n\leq120$ reference by only about
$0.003\%$. As the dipole is moved away from the surface, the
high-order contributions decay rapidly and the bound becomes
progressively concentrated in the lowest spherical channels.

Figure~\ref{fig:finite_source_bound}(c) shows the corresponding
fractional modal contributions
$\mathcal B_n/\sum_m\mathcal B_m$ for $n=1,\ldots,20$; the
denominator contains the converged modal sum. The logarithmic color
scale resolves the weaker high-order contributions. At $g=0.1$ the
largest individual contribution occurs near $n=10$, whereas near
$g=1$ the dipolar channel is dominant. The figure therefore gives a
direct visualization of how the contributions to the source-weighted broadband
absorption bound are distributed over angular momentum.

\section{Conclusions}
\label{sec:conclusion}

We have derived broadband absorption constraints for the
electromagnetic spherical modes of a material-coated PEC sphere.
Starting from the exact TE and TM modal reflection coefficients,
their long-wavelength asymptotics, together with causality and
contour integration in the complex wavelength plane, yield an
exact sum rule relating the weighted logarithmic reflection
integral to the static material properties and to the complex
reflection zeros. The spectral weight $\lambda^{2n}$ is set by
the long-wavelength behavior of the $n$th spherical channel and
therefore leads naturally to an angular-momentum-dependent
hierarchy of absorption--bandwidth constraints.

For the dipolar channels, $n=1$, the contribution of the
reflection zeros can be bounded analytically using passive
spherical-wave sum rules, establishing a rigorous
material-dependent broadband absorption bound for both TE and
TM polarizations. For $n\geq2$, the corresponding result depends
on a nonnegativity condition involving higher odd moments of the
reflection zeros, formulated here as Conjecture~1. Extensive
numerical tests over passive dispersive material models found no
counterexample and include realizations approaching the proposed
boundary, but a general analytical proof remains open. The
higher-order zero-moment condition therefore constitutes a
specific mathematical problem whose resolution would complete
the modal theory.

The bounds also expose the geometrical role of curvature. The
factor $(a/b)^{2n+1}$ governing the quasistatic coupling between
the coating interfaces originates from the radial dependence of
the $n$th spherical multipole. In the locally planar limit,
$b,n\rightarrow\infty$ with
$k_t\simeq(n+1/2)/b$ fixed and $d=b-a$ finite, this factor tends
to $e^{-2k_td}$. Thus, angular momentum on the sphere plays the
role of transverse momentum in a planar geometry, and curvature
transforms the conventional thickness--bandwidth limitation into
a discrete family of transverse-wavenumber-resolved constraints.

Finally, we used this modal viewpoint to bound absorption from a
localized source outside the sphere. A radial electric dipole on
the symmetry axis excites only the $m=0$ TM channels, allowing
its near field to be decomposed directly into the same spherical
modes for which the bounds were derived. The resulting
source-weighted bound is obtained by combining the individual
modal bounds according to the source excitation coefficients.
For a source close to the surface, the bound receives substantial
contributions from a broad range of spherical orders, whereas
increasing the source--surface separation suppresses the
higher-order components and concentrates the response in the
lowest channels. This finite-source example gives a direct
physical interpretation of the correspondence
$k_t\simeq(n+1/2)/b$: spatial localization of the incident near
field is translated into excitation of higher angular momenta,
each of which remains subject to its own causal broadband
absorption constraint.

\section*{Acknowledgment}
This work was supported in part by the Israel Science Foundation under Grant 1457/23, and by the Binational US-Israel Science Foundation under Grant 2024281.

\appendices

\renewcommand{\theequation}{\thesection\arabic{equation}}
\renewcommand{\theHequation}{appendix.\thesection.\arabic{equation}}

\section{Riccati--Bessel Small Argument Forms}
\label{app:hankel_expansions}
\setcounter{equation}{0}

For $x\ll1$, the Riccati--Bessel functions $\hat H_n^{(1,2)}(x)=x\,h_n^{(1,2)}(x)$ have the small-argument forms \cite{harrington,balanis}
{
\begin{equation}
\begin{aligned}
\hat H_n^{(1,2)}(x)
&\approx
\frac{x^{n+1}}{(2n+1)!!}
\mp j(2n-1)!!\,x^{-n},
\\
\hat H_n^{(1,2)\prime}(x)
&\approx
\frac{(n+1)\,x^{n}}{(2n+1)!!}
\pm jn(2n-1)!!\,x^{-(n+1)},
\end{aligned}
\label{eq:hankel_expansions}
\end{equation}
}%
from which Eqs.~\eqref{eq:R_full_asymptotic_n}--\eqref{eq:Cm} follow.

\section{All-Pass Arc Contribution}
\label{app:all_pass}

\setcounter{equation}{0}

The all-pass factor is defined in Eq.~\eqref{eq:HAP}.
For large $|\lambda|$, using
$\ln(1-z)=-\sum_{k=1}^{\infty}z^k/k$, we obtain
\begin{equation}
\begin{aligned}
\ln H_{\mathrm{AP}}(\lambda)
&=
\sum_{i=1}^{N}
\left[
\ln\left(1-\frac{\lambda_i^*}{\lambda}\right)
-
\ln\left(1-\frac{\lambda_i}{\lambda}\right)
\right]
\\
&=
2j\sum_{k=1}^{\infty}
\frac{1}{k\lambda^k}
\sum_{i=1}^{N}
\Im\left\{\lambda_i^k\right\}.
\end{aligned}
\end{equation}
Multiplying by $W(\lambda)=\lambda^{2n}$ gives
\begin{equation}
\begin{aligned}
\int_{C_{1/2}^{\Lambda}}
\lambda^{2n}\ln H_{\mathrm{AP}}(\lambda)\,d\lambda
&=
2j\sum_{k=1}^{\infty}\frac{1}{k}
\sum_{i=1}^{N}\Im\!\left\{\lambda_i^k\right\}
\\
&\qquad\times
\int_{C_{1/2}^{\Lambda}}
\lambda^{2n-k}\,d\lambda.
\end{aligned}
\end{equation}
For $k<2n+1$, the corresponding terms are purely imaginary and do not contribute to the real part. terms with
$k>2n+1$ vanish as $\Lambda\rightarrow\infty$. Thus, the finite term entering the
bound is obtained for $k=2n+1$.
Using
\begin{equation}
\lambda=\Lambda e^{j\phi},
\qquad
0\leq\phi\leq\pi,
\qquad
\int_{C_{1/2}^{\Lambda}}\frac{d\lambda}{\lambda}=j\pi\,
\end{equation}
the all-pass contribution becomes
\begin{equation}
\begin{aligned}
&\lim_{\Lambda\to\infty}
\Re\!\int_{C_{1/2}^{\Lambda}}
\lambda^{2n}\ln H_{\mathrm{AP}}(\lambda)\,d\lambda
\\
&\qquad =
\Re\!\left\{
\frac{2j^2\pi}{2n+1}
\sum_{i=1}^{N}
\Im\!\left\{\lambda_i^{2n+1}\right\}
\right\}
\\
&\qquad =
-\frac{2\pi}{2n+1}
\sum_{i=1}^{N}
\Im\!\left\{\lambda_i^{2n+1}\right\}.
\end{aligned}
\end{equation}
which reproduces Eq.~\eqref{eq:HAP_arc}.

\section{Proof of the Dipolar-Mode Bound (Theorem~\ref{thm:dipole})}
\label{app:dipole_proof}

\setcounter{equation}{0}

For the dipolar modes, $n=1$, the nonnegativity
condition in (\ref{eq:nonnegative_combined_higher_modes}) can be established analytically
from passivity. In this case, Eq.~(\ref{eq:nonnegative_combined_higher_modes}) becomes
\begin{equation}
\pi A_1+\frac{2\pi}{3}
\sum_i \Im\left\{\lambda_i^3\right\}\geq 0 ,
\label{eq:n1_condition}
\end{equation}
where, from Eq.~(\ref{eq:An}),
\begin{equation}
A_1=\frac{2(2\pi b)^3}{3}.
\label{eq:A1}
\end{equation}

We first recall a passive-system sum rule for electromagnetic
vector spherical waves. For a passive scatterer contained within
a sphere of radius $b$, the diagonal spherical-wave reflection
coefficient satisfies \cite{Bernland2011VSW,Bernland2011SumRules}
\begin{equation}
\frac{1}{\pi}\int_0^\infty
\frac{\ln |S_{\nu\nu}(k)|^{-1}}{k^2}\,dk
=
b-\frac{\beta_{\nu\nu}}{2}
+\sum_i\Im\left\{\frac{1}{\kappa_i}\right\},
\label{eq:Bernland_first_sum_rule}
\end{equation}
where $\kappa_i$ denote the reflection zeros in the causal
upper half of the complex wavenumber plane and
$\beta_{\nu\nu}\geq0$. The phase factor $e^{2jkb}$ associated with propagation
to the circumscribing sphere has unit magnitude on the real
frequency axis and is zero-free, so it does not affect the integral or the
locations of the reflection zeros.

To relate (\ref{eq:Bernland_first_sum_rule}) to the wavelength-plane
convention used here, recall that the present work assumes an
$\exp(j\omega t)$ time dependence and places the relevant reflection
zeros $\lambda_i$ in the upper half of the complex $\lambda$ plane.
The two conventions are related by complex conjugation, under which a zero $\lambda_i$ corresponds to a wavenumber-plane zero $\kappa_i=2\pi/\lambda_i^*$ of Ref.~\cite{Bernland2011VSW}; equivalently, since the zero set of a real passive response is symmetric under $\lambda\to-\lambda^*$, one may use $\kappa_i=-2\pi/\lambda_i$. Either way,
\begin{equation}
\Im\left\{\frac{1}{\kappa_i}\right\}
=
-\frac{\Im\{\lambda_i\}}{2\pi}.
\end{equation}
Since the integrand on the left-hand side of
(\ref{eq:Bernland_first_sum_rule}) is nonnegative for a passive
system, and $\beta_{\nu\nu}\geq0$, it follows that
\begin{equation}
\sum_i\frac{\Im\{\lambda_i\}}{2\pi}
\leq b-\frac{\beta_{\nu\nu}}{2}
\leq b.
\label{eq:first_zero_bound}
\end{equation}

We now define the normalized zero locations
\begin{equation}
z_i=\frac{\lambda_i}{2\pi b}=x_i+jy_i,
\qquad y_i>0.
\label{eq:normalized_zeros}
\end{equation}
Equation~(\ref{eq:first_zero_bound}) gives
\begin{equation}
\sum_i y_i\leq 1.
\label{eq:y_sum}
\end{equation}
For each zero,
\begin{equation}
\Im\{z_i^3\}
=
3x_i^2y_i-y_i^3
\geq -y_i^3.
\label{eq:cubic_zero}
\end{equation}
Therefore,
\begin{align}
\sum_i\Im\{z_i^3\}
&\geq-\sum_i y_i^3 \nonumber\\
&\geq-\left(\sum_i y_i\right)^3
\geq-1,
\label{eq:cubic_sum_bound}
\end{align}
where the second inequality follows from $y_i\geq0$ and the last
one from (\ref{eq:y_sum}). Since $\lambda_i=2\pi b z_i$,
(\ref{eq:cubic_sum_bound}) yields
\begin{equation}
\sum_i\Im\{\lambda_i^3\}
\geq -(2\pi)^3b^3.
\label{eq:lambda_cubic_bound}
\end{equation}

Finally, substituting (\ref{eq:A1}) and
(\ref{eq:lambda_cubic_bound}) into
(\ref{eq:n1_condition}) gives
\begin{align}
\pi A_1+\frac{2\pi}{3}\sum_i\Im\{\lambda_i^3\}
&\geq
\frac{2\pi}{3}(2\pi b)^3
-\frac{2\pi}{3}(2\pi)^3b^3 =0.
\end{align}
Hence,
\begin{equation}
\pi A_1+\frac{2\pi}{3}
\sum_i\Im\{\lambda_i^3\}\geq0
\label{eq:n1_proven}
\end{equation}
for both the TE and TM dipolar channels, which proves Theorem~\ref{thm:dipole}. Note that the argument uses only passivity and the circumscribing radius $b$, so the dipolar bound holds for any passive scatterer contained in the sphere of radius $b$, not only for the coated PEC sphere.

For $n\geq2$, the corresponding condition involves the higher odd moments $\Im\{\lambda_i^{2n+1}\}$, and the pointwise inequality \eqref{eq:cubic_zero} has no analogue of the required sign (Section~\ref{sec:bounds}); the higher-order conditions are therefore examined numerically in Section~\ref{sec:numerics} and posed as Conjecture~\ref{conj:zeros}, left as an open challenge for the community.

\balance
\bibliographystyle{IEEEtran}
\bibliography{references}
\end{document}